\documentstyle[aps,prb,multicol,epsfig]{revtex}
\begin{document}
\title{Is doped BaBiO$_3$ a conventional superconductor?}
\author{V. Meregalli and S. Y. Savrasov }
\address{Max-Planck-Institut f\"ur Festk\"orperforschung,
Heisenbergstr. 1, 70569 Stuttgart, Germany.}
\date{\today }
\maketitle

\begin{abstract}
We report density functional calculations based on local density
approximation (LDA) of the properties of doped barium bismuthates. Using
linear-response approach developed in the framework of the linear
muffin-tin-orbital method the phonon spectrum of the Ba$_{0.6}$K$_{0.4}$BiO$%
_3$ system is calculated and is compared with the results of the neutron
diffraction measurements. The effect of doping in the calculation is
modelled by the virtual crystal and mass approximations. The electron-phonon
coupling constant $\lambda $ is then evaluated for a grid of phonon
wave-vectors using the change in the potential due to phonon distortion found
self-consistently. A large coupling of the electrons to the bond-stretching
oxygen vibrations and especially to the breathing-like vibrations is
established. Also, a strongly anharmonic potential well is found for
tilting-like motions of the oxygen octahedra. This mode is not coupled to
the electrons to linear order in the displacements, therefore an anharmonic
contribution to $\lambda $ is estimated using frozen--phonon method. Our
total (harmonic plus anharmonic) $\lambda $ is found to be 0.34. This is too
small to explain high-temperature superconductivity in Ba$_{0.6}$K$_{0.4}$BiO%
$_3$ within the standard mechanism. Finally, based on standard LDA and LDA+U
like calculations, a number of properties of pure BaBiO$_3$ such as tilting
of the octahedra, breathing distortion, charge disproportionation and
semiconducting energy gap value is evaluated and discussed in connection
with the negative $U$ extended Hubbard model frequently applied to this
compound.
\end{abstract}


\begin{multicols}{2}

\begin{center}
{\bf I. INTRODUCTION.}
\end{center}

Since the discovery of superconductivity at $T_c\sim $30K in Ba$_{1-x}$K$_x$%
BiO$_3$ (BKBO)\ \cite{DISC,REVIEW}, and from earlier studies of BaPb$_{1-x}$%
Bi$_x$O$_3$ (BPBO) system with $T_c\sim $13K, there is a fundamental
question whether the conventional phonon mediated pairing mechanism is
operative in these high--$T_c$ superconductors (HTSC). Doped barium
bismuthates are different from the HTSC cuprates  \cite{Plakida}, since no
antiferromagnetic ordering exists for the parent compound BaBiO$_3.$ This
seriously doubts that strong electron correlations exist and are responsible
for the pairing. Simple cubic superconducting phase makes BKBO and BPBO
similar to the isotropic low $T_c$ superconductors. However, there is a
number of features which makes doped bismuthates similar to the cuprates.
Both systems are perovskite oxide superconductors with surprisingly low
density of states at the Fermi level. This can hardly give high transition
temperatures for the BCS--like superconductors. As high--$T_c$ cuprates are
originated from antiferromagnetic insulators, the parent BaBiO$_3$ compound
is a charge--density wave (CDW) insulator in which oxygen octahedra around
the Bi ions exhibit alternating breathing--in and breathing--out distortions.
The Bi ions exist in the charge disproportionate state which is chemically
interpreted as 2Bi$^{4+}\Rightarrow $Bi$^{3+}$+Bi$^{5+}$. It is therefore
tempting to connect the mechanism of superconductivity with the nature of
these insulators.

Unfortunately, experimental estimates of the electron--phonon coupling
strength do not lead to the firm conclusion on the origin of
superconductivity in the bismuthates. Large isotope effect with $\alpha $%
=0.4 has been reported for BKBO \cite{ISO2}. Other measurements \cite{ISO1}
give $\alpha $=0.21$\pm $0.03 for BKBO and $\alpha $=0.22$\pm $0.03 for
BPBO. Using their analysis, the authors of Ref. \onlinecite{ISO1} concluded
that ''phononic'' effects in these materials are only indicative of dressed
electronic excitations. From studying the imaginary part of optical
conductivity in BKBO the authors of Ref. \onlinecite{OPT} gave the value of $%
\lambda \sim $0.2. Electronic specific heat measurements \cite{HEAT} of Ba$%
_{0.6}$K$_{0.4}$BiO$_3$ have yielded $N_s^{*}(0)\sim $0.32 states/[spin$%
\times $eV$\times $cell] giving a mass enhancement factor $%
N_s^{*}(0)/N_s^{band}(0)\sim $1.4 ($N_s^{band}(0)\sim $0.23 states/[spin$%
\times $eV$\times $cell] for x=0.4). A complicated situation exists with the
transport measurements. The temperature dependent resistivities for
superconducting BKBO and BPBO have ranged from metallic to semiconducting
and two--channel model of the conductivity in the bismuthates was discussed 
\cite{RES}. While good grain--boundary--free thin films and single crystals
of BKBO doped well away from the CDW instability seem to exhibit metallic
behaviour, the values of the resistivity itself are (like in the cuprates)
unusually high and are of the order a few hundred $\mu \Omega \times $cm at
room temperature This could point out that an additional (to standard
electron--phonon) scattering mechanism is presented.

The most direct evidence on the importance of electron--phonon interactions
in superconductivity of the bismuthates has been given by the tunnelling
measurements \cite{TUN1,TUN2}. Although not identical for different
junctions, the deduced Eliashberg spectral functions $\alpha ^2F(\omega )$
bear a close resemblance with the phonon density of states determined by
inelastic neutron scattering \cite{NEUT}. The estimated values of $\lambda
\, $ vary from 0.7 to 1.2 which seem to be sufficient to explain high
critical temperatures\ within the standard mechanism.

The electron--phonon coupling in doped BaBiO$_3$ has been investigated
theoretically by several methods. The authors of Ref. \onlinecite{TB} study
this problem using tight--binding (TB) fit to the energy bands which are
obtained from density--functional calculations based on local--density
approximation \cite{LDA} (LDA). The computed Eliashberg spectral function $%
\alpha ^2F(\omega )$ has been found to display prominent features in the
frequency range corresponding to the oxygen stretching modes and the value
of $\lambda $=1.09 has been reported. Crude calculations based on
rigid--muffin--tin approximation (RMTA) also give large $\lambda $ $\sim $3
indicating a strong--coupling regime \cite{Freem}. Two estimates of $\lambda 
$ using total--energy frozen--phonon method appeared in the literature \cite
{Liecht1,Kunc}. Note that, in contrast to the TB and RMTA\ methods, the
frozen--phonon calculations treat screening of the potential due to lattice
distortions self--consistently. The value of the electron--phonon coupling
strength equal to 0.3 for the breathing mode has been found \cite{Liecht1}
and the rough estimate of $\lambda \sim $0.5 was obtained \cite{Kunc} using
12 ${\bf q}$=0 phonons for ordered cubic Ba$_{0.5}$K$_{0.5}$BiO$_3$.

There was a partial success in predicting structural phase diagram for the
parent compound BaBiO$_3$ within the LDA \cite{Liecht1,Zeyher,Blaha,Liecht2}%
. The experimental structure mainly consists of combined tilting and
breathing distortions of the oxygen octahedra corresponding to the instable
R--point phonons of the cubic phase \cite{STRUC,STRUC2}. While rotational
instability was found in all calculations, the frozen--in breathing mode was
not described by pseudopotential calculation \cite{Zeyher} and two
linear--muffin--tin--orbital (LMTO) calculations \cite{Liecht1,Liecht2} give
the value for the breathing distortion about 30\% off the experimental one.
Less rigorous potential--induced--breathing (PIB) model obtained both
instabilities \cite{PIB} with similar accuracy. It is not clear whether
these discrepancies are due to sensitivity to computational details or due
to the local density approximation itself.

A great amount of work \cite{REVIEW} has been done to understand the
properties of the barium bismuthates on the basis of the negative $U$
extended Hubbard model originally introduced in Ref. \onlinecite{Anderson}.
The valence configuration of semiconducting BaBiO$_3$ can be viewed as Ba$_2$%
Bi$^{3+}$Bi$^{5+}$O$_6$ which represents a lattice of electron pairs
centred at every second Bi site (Bi$^{3+}$). The sites occupied with Bi$%
^{5+}\,$ions are interpreted as those with no electrons. Rice and
co--workers \cite{Rice} have proposed that such local pairs are stabilised
by polarising the O octahedra and the effective on--site $U$ becomes
negative due to the large electron--phonon coupling. Recently, Varma \cite
{Varma} has pointed out that negative $U$ can also be of the electronic
origin due to the skipping of the valence ''4+'' by the Bi ion. The latter
can provide a possible explanation for a well-separated optical and
transport energy gap in the bismuthates\cite{Uchida}. The mean--field phase
diagram of the negative $U$ Hubbard model exhibits several stable phases
involving a CDW semiconductor, and a singlet superconductor. This is in
qualitative agreement with the experimental phase diagram \cite{Taraphder}.

The question on the origin of negative $U$ is of great interest since it may
provide an insight on the superconductivity mechanism in the bismuthates.
Recent calculations using constrained density--functional theory have been
carried out to obtain the Coulomb interaction parameters for the Bi $6s$
orbitals \cite{Vielsack}. No indication for negative $U$ of the electronic
origin was reported.

In the present work we try to address several problems seen from the above
introduction by means of the state--of--the-art density functional LDA
calculations. As a first problem, we study lattice dynamics of the
superconducting cubic Ba$_{0.6}$K$_{0.4}$BiO$_3$ using recently developed
linear-response approach implemented within the LMTO method\cite{PHN}. The
effect of doping is modelled by the virtual crystal and virtual mass
approximations. On the basis of this calculation, we estimate
electron--phonon coupling in this compound. The linear--response method used
by us is advantageous in contrast to the frozen--phonon approach since it
allows the treatment of perturbations with arbitrary wave vectors ${\bf q}$.
We have demonstrated its accuracy by calculating lattice--dynamical,
superconducting and transport properties for a large variety of metals \cite
{EPI}, and we believe that our calculated value of $\lambda $ will be a
realistic estimate for the electron--phonon coupling strength in this high--$%
T_c\,$superconductor. Also, a recent publication \cite{ILPRL} deals with the
application of the linear--response method to study the electron--phonon
interaction in another high--$T_c$ superconductor CaCuO$_2$.

The second problem, we focus in our work, is studying the effects of
anharmonicity in the electron-phonon coupling. It is widely accepted that
certain phonon modes are strongly anharmonic in the high--$T_c$ materials.
Frozen--phonon calculations produce double--well potentials for buckling
motions of oxygen atoms perpendicular to the CuO planes in nearly all HTSC 
\cite{ILPRL,Rodriguez,Pickett}, chain--buckling distortions are found to be anharmonic 
\cite{Pickett} in YBa$_2$Cu$_3$O$_7$, X--point tiltings of the octahedra
along (110) directions are instable \cite{Krak} in La$_2$CuO$_4$, and
R--point instabilities corresponding to breathing and tilting exist in the
doped barium bismuthates \cite{Liecht1,Liecht2,Kunc}. The influence of
anharmonicity to high--temperature superconductivity has been addressed in
several publications \cite{Hardy,Schuttler,Cohen}, especially because of the
small isotope effect found for HTSC cuprates. Triple--degenerate pure
rotational mode at the R point of cubic ordered Ba$_{0.5}$K$_{0.5}$BiO$_3$
was predicted to exhibit a double--well potential and some estimates of the
anharmonic contributions to $\lambda $ have been given \cite{Kunc}. We
extend this analysis by solving numerically Schr\"{o}dinger's equation for
the anharmonic potential well found from frozen--phonon calculations. The
anharmonic $\lambda $ is then computed along the lines proposed in Refs. %
\onlinecite{Cohen,Hui} by estimating the electron--phonon matrix elements
from the energy bands computed for different tilting distortions. We
conclude, in accord with the previous findings \cite{Kunc}, that this
contribution, while not decisive, is not negligible for the total value of $%
\lambda $.

The third purpose of our work is to study the properties of the undoped
parent compound BaBiO$_3$. We try to answer the question whether the LDA
gives an adequate description of the ground state properties for this
charge--density--wave insulator. Since there was some inconsistency reported
in previous calculations \cite{Liecht1,Kunc,Zeyher,Blaha,Liecht2}, we want
to rule out possible sensitivity of the final results to the internal
parameters used in our band structure calculations with the full--potential
LMTO method \cite{FPLMTO}. We carefully choose our LMTO basis set, number of 
${\bf k}$ points, plane wave energy cutoff and other parameters by examining
the convergency of the total energy and the calculated properties with
respect to them. Based on the well converged data, we come to the conclusion
that the breathing distortions are {\em seriously underestimated} (ideally,
absent) in the LDA, and, therefore, the insulated state is not correctly
described. This strongly resembles the situation with the antiferromagnetic
ground state of the cuprates superconductors which is also not described by
the LDA \cite{Pick}. We perform a number of model calculations in the spirit
of the LDA+U method \cite{LDA+U} in order to clarify this problem.

The rest of the paper is organised as follows: In Sec. II, our
linear--response calculations of the lattice dynamics and the
electron--phonon interactions in doped BaBiO$_3$ are described. Sec. III
considers anharmonicity corrections to $\lambda $ for the tilting motions of
the oxygen octahedra. Sec. IV reports our LDA and model calculations of the
ground state properties for pure BaBiO$_3.$ In Sec. IV we give our
conclusions.

\begin{center}
{\bf II. HARMONIC\ PHONONS\ AND\ }${\bf \lambda }$
\end{center}

This section presents our results on the lattice dynamics and the
electron--phonon interaction for the cubic perovskite superconductor Ba$%
_{0.6}$K$_{0.4}$BiO$_3.$ We also summarise the main features of the
calculated electronic structure and discuss our predicted equilibrium
lattice configuration. The band structure calculations are performed with
the highly precise full--potential LMTO method \cite{FPLMTO}. The details
of the calculations are the following: The effect of potassium doping is
taking into account using virtual crystal approximation (VCA) by considering
a fractional nuclei charge $Z$=55.6 at the Ba site. Numerous supercell
investigations \cite{Kunc,Mattheiss} of the doping influence on the
calculated energy bands justify the applicability of the VCA. A multiple,
three--$\kappa $ LMTO basis set with the tail energies equal to -0.1,-0.8,
and -2 Ry. is employed for representing valence wave functions. The valence
states include $6s$ and $6p$ orbitals of Bi, $2p$ orbitals of O, and $6s$
orbitals of Ba. Such semicore states as $5d$ orbitals of Bi, $2s$ orbitals
of O, and $5p$ orbitals of Ba are treated as bands and are included in the
main valence panel using two $\kappa $ LMTO basis with $\kappa _{1,2}^2$%
=-0.1,-0.8 Ry. The main panel also includes unoccupied $5d$ orbitals of Ba
with the 2$\kappa $ basis and $4f$ orbitals of Ba with the 1$\kappa $ basis (%
$\kappa ^2$=-0.1 Ry). Deeper lying $5s$ states of Ba are resolved in a
separated energy panel. All other states are treated as core levels. The
muffin--tin sphere radii were taken to be: $S_{\text{Ba}}$=3.25 a.u., $S_{%
\text{Bi}}$=2.25 a.u., and $S_{\text{O}}$=1.80 a.u. All calculations are
performed at the experimental lattice constant a=8.10 a.u. The
Barth--Hedin--like exchange--correlation formula after Ref. %
\onlinecite{Moruzzi} is used. The valence bands are treated scalar
relativistically and the core levels - fully relativistically. A number of $%
{\bf k}$ points for the Brillouin zone (BZ) integration using an improved
tetrahedron method \cite{TET} is taken to be 20 per$\frac 1{48}$th BZ. The
charge density and the potential in the interstitial region are expanded in
plane waves with the cutoff corresponding to the (28,28,28)
fast--Fourier--transform (FFT) grid in the real space (approximately 10000
plane waves).

\begin{figure}[tb]
\begin{minipage}[tb]{3.25in}
\begin{center}
\epsfig{file=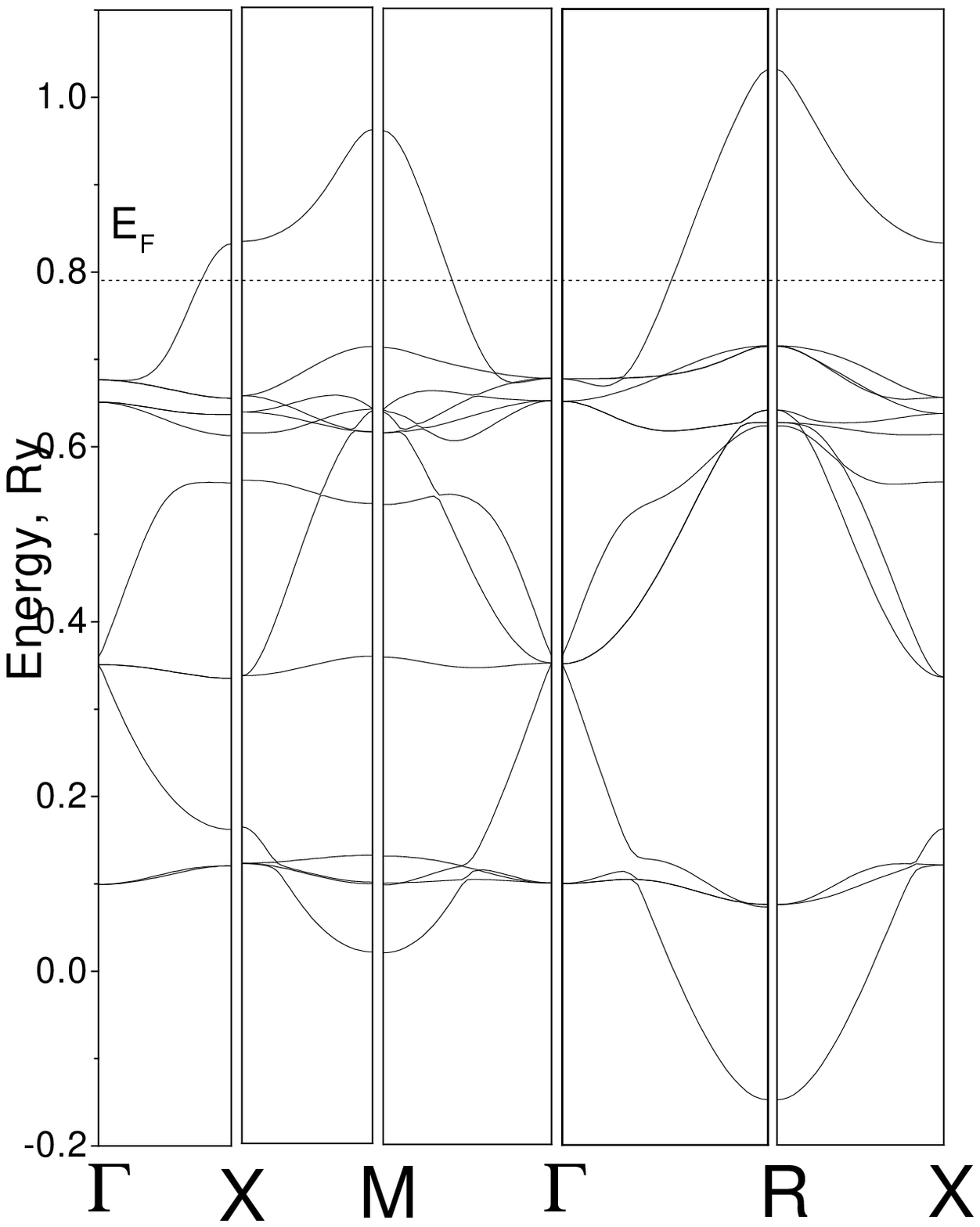,height=2.0in,width=2.5in,clip=}
\end{center}
\caption{Calculated LMTO energy bands for cubic Ba$_{0.6}$K$_{0.4}$BiO$_3$.
The potassium doping is taken
into account using virtual crystal approximation.}
\label{bands}
\end{minipage}
\end{figure}

We first summarise the main features of the calculated electronic structure
in doped BaBiO$_3$. The occupied part of the bands (see Fig. \ref{bands})
mainly consists of Bi$(6s)$--O$(2p)$ hybridised band complex. This is in
accord with the previous calculation\cite{Mattheiss}. For the cubic
perovskite phase, there is only one band crossing the Fermi level, which is
an antibonding Bi--O $sp(\sigma )$ band. Similar situation is found in the
cuprate superconductors where Cu--O $dp(\sigma )$ antibonding bands dominate
at the Fermi energy $E_F$. A simple tight--binding model involving Bi$(6s)$,
O$(2p)$ orbitals, and two--centre nearest--neighbour $sp(\sigma )$
interaction can be used to understand the principal features of these energy
bands\cite{MatHam}. It was early noted \cite{MatHam} that for the case of
half--filling (undoped cubic BaBiO$_3$) this model has a perfectly nested
Fermi surface for the wave vector ${\bf q}$ corresponding to the R--point.
Therefore, it is tempting to interpret the appearance of breathing
distortions as commensurate Peierls instability and cubic perovskite Ba$%
_{0.6}$K$_{0.4}$BiO$_3$ as a doped Peierls insulator \cite{Rice2}. To
understand whether nesting can bring any effect in static susceptibility, we
have analysed ${\bf q}$--dependence of the function

\begin{equation}
\sum_{{\bf k}jj^{\prime }}\delta (E_{{\bf k}j})\delta (E_{{\bf k}+{\bf q}%
j^{\prime }})  \label{f1}
\end{equation}

\begin{figure}[bt]
\begin{minipage}[t]{3.25in}
\begin{center}
\epsfig{file=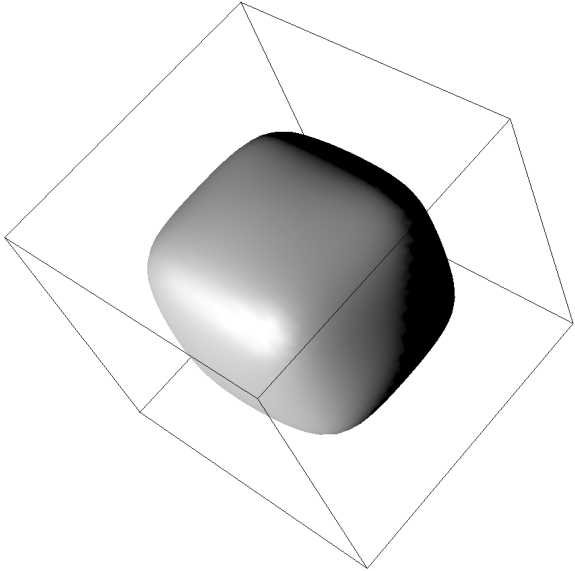,height=3.0in,width=3.0in,clip=}
\end{center}
\caption{Calculated Fermi surface for cubic Ba$_{0.6}$K$_{0.4}$ BiO$_3$.
using the LMTO method. The effect of potassium doping is taken
into account within the virtual crystal approximation. The centre of the cube
corresponds to the $\Gamma $ point of the Brillouin zone.}
\label{fs}
\end{minipage}
\end{figure}

for the realistic energy bands $E_{{\bf k}j}$ (relative $E_F$) using
experimental structures. We conclude that the nesting is far from perfect in
the case of half--filling and dimerisation of the oxygen octahedra can
hardly be connected with it. A realistic TB model should also include Bi$%
(6p) $ orbitals and their nearest--neighbour interaction with O$(2p)$ states 
\cite{MatHam,Vielsack}. Upon potassium doping, the bands hardly change
except for a slight lowering of $E_F\,$ away from half--filling. For Ba$%
_{0.6}$K$_{0.4}$BiO$_3$ the Fermi surface represents a rounded cube centred
at the $\Gamma \,\,$point as shown in Fig. \ref{fs}. Analysis of the band
structure factor given by Eq. (\ref{f1}) as a function of ${\bf q}$ shows
featureless behaviour and any effect of the nesting enhancement on the
electron--phonon interaction is not expected for this band dispersion.

We second discuss our results for the calculated equilibrium lattice
configuration in Ba$_{0.6}$K$_{0.4}$BiO$_3$. The
theoretical--to--experimental volume ratio $V/V_{exp}$ is found to be 1.01,
and the calculated bulk modulus is equal to 1.25 Mbar. Both neutron
diffraction\cite{STRUC2} and x--ray--absorption--fine--structure\cite{XAFS}
(XAFS)\ measurements show that frozen--in breathing distortions are absent
in the superconducting phase. We have performed frozen--phonon calculations
for the doubled cell corresponding to the $R$--point and for several
breathing distortions. The total energy minimum shows that the undistorted
cubic phase is stable in accord with these experiments. The curvature is
well fitted with standard parabola, which shows that the breathing mode is
harmonic in the superconducting phase.

We further investigate tilting of the octahedra. Experimentally, for the
undoped compound the octahedra rotated\cite{STRUC2} at the angle $\sim $11.2$%
^{\circ }$ along (1,1,0) axe. More complicated situation exists in the
superconducting phase. According to the neutron diffraction data \cite{STRUC}%
, the average structure is cubic although presence of a weak long--range
superstructure characterised by the octahedra rotations at the angles about 3%
$^{\circ }$ was also found\cite{Braden,Braden2}. Recent XAFS measurements 
\cite{Jacoby} report on the locally disordered rotations. From their
analysis, the authors of Ref. \onlinecite{Jacoby} conclude that the
rotations can either be along (1,1,1) or (1,1,0) axe. Previous
frozen--phonon calculations \cite{Kunc} performed for the ordered Ba$_{0.5}$K%
$_{0.5}$BiO$_3$ investigate (1,0,0) component of the tilting mode which is
found to be unstable with the total energy minimum corresponding to the
angle 7$^{\circ }$.

\begin{figure}[bt]
\begin{minipage}[t]{3.25in}
\begin{center}
\epsfig{file=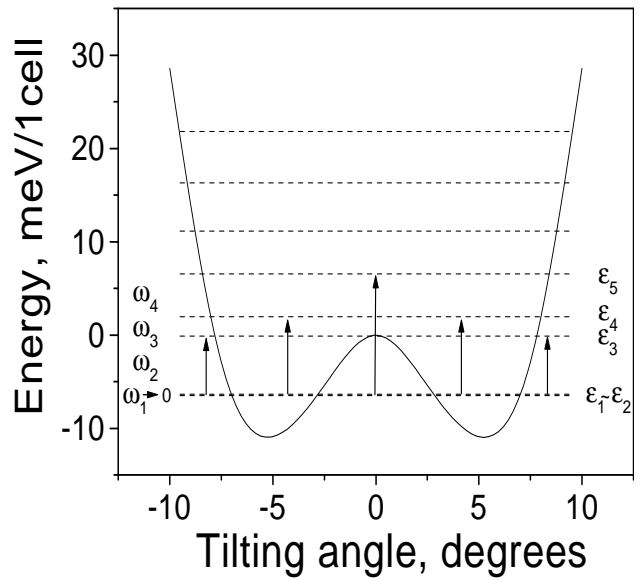,height=3.0in,width=\linewidth,clip=}
\end{center}
\caption{Frozen--phonon calculation of the total energy
(meV/1$\times $cell) as a function of the tilting angle in 
Ba$_{0.6}$K$_{0.4}$BiO$_3$. The levels $\epsilon _n$ are
the solutions of the Schr\"odinger equation for the 
anharmonic oscillator with the double--potential well 
shown on the figure. The transitions 
$\omega _n=\epsilon _n-\epsilon _0$ involving different
phonon excited states are illustrated by arrows.}
\label{tilt}
\end{minipage}
\end{figure}

Our own total--energy calculations also confirm the existence of tilting
distortions. Fig. \ref{tilt} shows that the total energy exhibits a double
well behaviour as a function of the rotation angle. We choose (1,1,0) axe for
the tilting as is the case in the undoped compound. The unit cell in the
calculation is doubled according to the R point of the cubic phase. The
total energy minimum is found at the angle equal to 5$^{\circ }.$ The energy
gain compared to the cubic phase is only 10 meV/(1$\times $cell) which
indicates that at the temperatures of the order $T_c$ the rotations can be
dynamic. The double well behaviour at such small energy scale unambiguously
points out on the importance of evaluating anharmonicity contribution in
total electron--phonon coupling. This problem will be discussed in the
following section.

We now report our main results on the calculated lattice dynamical
properties of Ba$_{0.6}$K$_{0.4}$BiO$_3$. The density functional
linear--response approach \cite{PHN} implemented on the basis of the
full--potential LMTO method \cite{FPLMTO} is used in this calculation. The
dynamical matrix is computed at 20 irreducible ${\bf q}$ points
corresponding to the (6,6,6) reciprocal lattice grid of the cubic BZ. The
effect of the potassium substitution on the phonon spectrum is taken into
account by virtual mass approximation. The LMTO basis set and other
technique details have been described above. One more comment should be said
on evaluating BZ integrals in the linear--response calculation. Here, one
can essentially improve the accuracy of the integration by using a multigrid
technique \cite{PHN}. A (6,6,6) grid (20 irreducible ${\bf k}$--points) is
used for finding linear--response functions while the effects of the energy
bands and the Fermi surface are taken into account using a (30,30,30) grid
(816 irreducible ${\bf k}$--points).

\begin{figure}[tbp]
\begin{minipage}[t]{6.5in}
\centerline{
\rotatebox{-90}{\resizebox{3.8in}{!}{\includegraphics{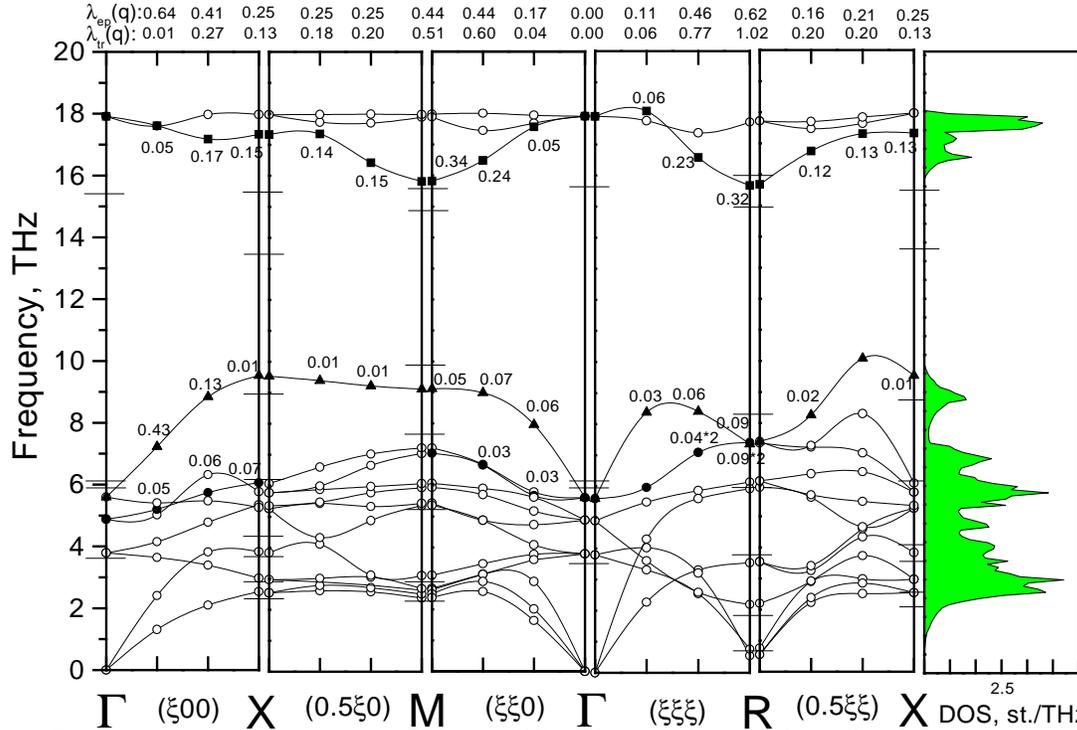}}}
}
\caption{Calculated phonon spectrum of Ba$_{0.6}$K$_{0.4}$BiO$_3$ using
density--functional linear--response method. The potassium doping
is taken into account using virtual crystal and virtual mass approximation.
The calculated points are shown by symbols.
The lines result from interpolation between the points. Horizontal bars
indicate the measured \cite{Braden2} phonon frequencies. Numbers for every phonon
mode indicate the
calculated electron--phonon coupling constants $\lambda _{{\bf q}\nu }$.
(Only the values larger than 0.01 are emphasised.)
On top of the figure shown are (i) the values of $\lambda $ summed over all
branches for given ${\bf q}$, (ii) the values of ${\bf q}$--dependent
transport constant $\lambda _{tr}$. The calculated phonon density of states 
$F(\omega )$ is shown on the right.}
\label{phn}
\end{minipage}
\end{figure}

The calculated phonon spectrum along major symmetry directions of the cubic
BZ is plotted in Fig. \ref{phn} . Solid circles denote the calculated points
and the lines result from interpolation between the circles. Several
features can be seen from these phonon dispersions. Three high--frequency
optical branches around the $\omega \sim $17 THz are well separated from the
other modes distributed in the frequency range from 0 to 10 THz. The
high--frequency modes mainly consist of the oxygen bond--stretching
vibrations. The longitudinal branch at the point $R$ corresponds to the
breathing mode which in our calculation has a frequency 15.7 THz. From the
analysis of our polarisation vectors, we conclude that oxygen bond--bending
vibrations dominate in the frequency interval between 6 and 10 THz. The
octahedra tilting modes are at the low--frequency interval. They exhibit
significant softening near the ${\bf q}$--point $R$=(1,1,1)$\pi /a$. Due to
symmetry, one can talk about pure tilting at the line between ${\bf q}$%
--point $M$=(1,1,0)$\pi /a$ and the $R$ point. Exactly at the $M$ point
nearest octahedra tilt in--phase and they tilt out--of--phase at the
R--point. A nearly--zero--frequency triple--degenerate mode exists at ${\bf q%
}$=(1,1,1)$\pi /a$ which corresponds to the pure rotational $T_{2u}$ phonon.
In fact, for $T$=0 this mode should have a slightly imaginary frequency for
the cubic structure according to our frozen--phonon analysis illustrated in
Fig. \ref{tilt} . But, due to numerical inaccuracies the linear--response
calculation gives very small positive $\omega $=0.5 THz. No significant
softening of the tilting modes near the point $M$ is predicted by our
calculation.

\end{multicols}

\begin{multicols}{2}

The phonon dispersion curves along $\Gamma X$, $\Gamma M,$ and $\Gamma R$
symmetry directions for Ba$_{0.6}$K$_{0.4}$BiO$_3\,\,$have been very
recently measured by inelastic neutron scattering \cite{Braden,Braden2}.
Horizontal lines in Fig. \ref{phn} indicate measured phonon frequencies at
the symmetry points $\Gamma $, $X$, $M,$ and $R$e as we were able to deduce
them from Fig. 1 of Ref. \onlinecite{Braden2}. The existence of soft
rotational modes near $R$ can be directly seen from the measured phonon
dispersions. The authors of Ref. \onlinecite{Braden2} have reported that
their samples still have a week long--range superstructure characterised by
the tilting of the octahedra. An extremely low frequency $\sim $0.9 THz of
these modes was measured. This is in agreement with our calculations.

Two other comments should be said on the comparison between our theory and
the experiment. One comment concerns frequency interval from 0 to 10 THz.
Here, our calculation is seen to reproduce measured phonon dispersions with
the accuracy of the order 10\%. In particular, the lowest mode in $\Gamma $
has $\omega _{calc}$=3.79 THz which can be compared with $\omega _{exp}\sim $%
3.5 THz. This mode mainly involves Ba(K) and Bi vibrations. The next mode in 
$\Gamma $ is the oxygen out--of--phase mode. Here, $\omega _{calc}$=4.88 THz
and the measured frequency is less than 6 THz. The so--called ferroelectric
mode has a frequency 5.59 THz in our calculation which is close to $\omega
_{exp}$ found near 6 THz. This mode is bond--bending longitudinal and it has
the strongest polar character. Usually it exhibits large splitting from the
TO mode at $\Gamma $ in cubic perovskites. The presence of free charge
carriers screens Coulomb interaction at long distances, and therefore, the
LO--TO splitting is absent in our calculation. The dispersion of the
ferroelectric mode as a function of ${\bf q}$ is also seen to be correctly
reproduced.

\begin{figure}[bt]
\begin{minipage}[t]{3.25in}
\begin{center}
\epsfig{file=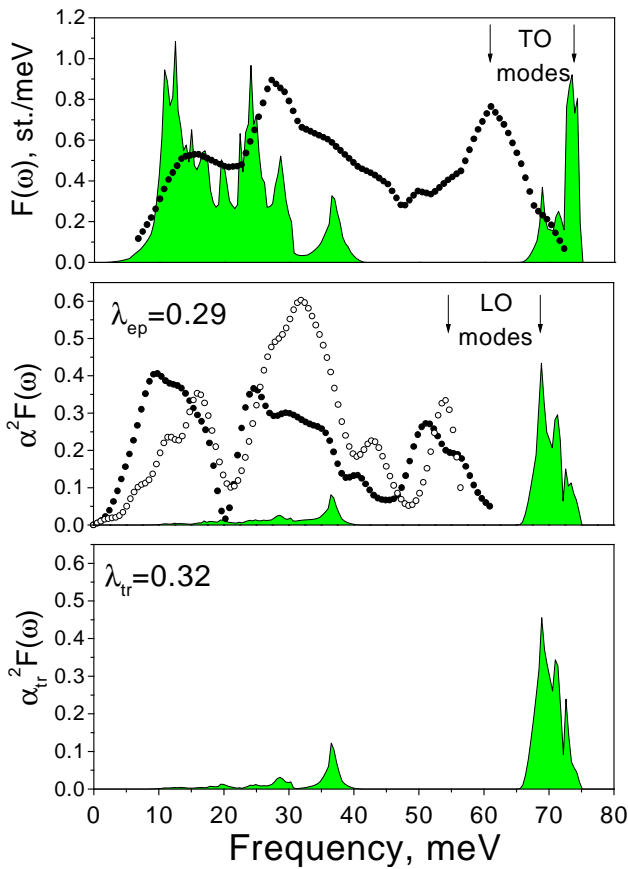,height=4.5in,width=\linewidth,clip=}
\end{center}
\caption{Results for doped BaBiO$_3$. (a) Comparison between calculated and
experimental \cite{NEUT}(symbols) phonon density of states. (b) Calculated Eliashberg
spectral function $\alpha ^2F(\omega )$ and the results of the
tunnelling measurements \cite{TUN1} (symbols). (c) Calculated transport
spectral function $\alpha {^{2}}{_{tr}}F(\omega )$.}
\label{dos}
\end{minipage}
\end{figure}

The second comment concerns our comparison for the high--frequency interval,
where the results of the calculations are found to be less accurate and the
overall discrepancy consists about 20\%. The highest mode at $\Gamma $ is
Bi--O bond--stretching mode. Here, we report the value of $\omega $ equal to
17.91 THz and the $\omega $ value only slightly larger than 15 THz was found
experimentally. The authors of Ref. \onlinecite{Braden} discuss an
anomalous dispersion for the longitudinal optical branch of the
one--dimensional breathing mode along $\Gamma X$ with its pronounced
frequency renormalisation. Our calculation, on the other hand, gives much
less dispersive optical branches along this direction as can be seen from
Fig. \ref{phn}. It is not clear whether this result is due to inhomogeneity
of the potassium distribution or other imperfectness of the samples used in
the experiment or due to drawbacks in our calculation connected with the
virtual crystal approximation. In fact, it is clearly seen that all our
high--frequency branches are overestimated by $\sim $20\% in contrast to the
experimental ones (except eventually the breathing vibrations near the point 
$R$). This result also follows from the comparison of our calculated and the
measured\cite{NEUT} phonon density of state $F(\omega )\,$[see Fig.\ref{dos}%
(a)]. It was found experimentally \cite{NEUT} that the oxygen
bond--stretching modes exhibit softening with the substitution of Ba by K.
These modes are located at the energies $\sim $70meV (or 17 THz) in undoped
BaBiO$_3$. Therefore it is tempting to connect possible source for the
discrepancies with our poor treatment of doping. The authors of Ref. %
\onlinecite{NEUT} discuss doping induced appearance of localised holes on
the oxygen 2p orbitals which screen the charge on the oxygen anions. This
charge reduction will lower the energy of these modes. If the localised hole
picture is correct, the VCA will not capture this since it removes electrons
from the conduction band by(uniform distributing the holes between O($2p$)
and Bi($6s$) orbitals.

We now report our results for the calculated electron--phonon interaction.
Based on our screened potentials which are induced by nuclei displacements
and are found self--consistently, we evaluate matrix elements of the
electron--phonon interaction, $g_{{\bf k}+{\bf q}j^{\prime }{\bf k}j}^{{\bf q%
}\nu }$. The standard expression \cite{comm1} for the electron--phonon
matrix elements reads as

\begin{equation}
g_{{\bf k}+{\bf q}j^{\prime }{\bf k}j}^{{\bf q}\nu }=\left\langle {\bf k}+%
{\bf q}j^{\prime }\left| \sum_{{\bf R}\alpha }\frac{Q_{{\bf R}\alpha }^{(%
{\bf q}\nu )}}{\sqrt{2M_{{\bf R}}\omega _{{\bf q}\nu }}}\delta _{{\bf R}%
\alpha }^{+}V\right| {\bf k}j\right\rangle  \label{f3}
\end{equation}
where $Q_{{\bf R}\alpha }^{({\bf q}\nu )}$ are the orthonormalised
polarisation vectors associated with the mode ${\bf q}\nu $, $M_R$ are the
atomic masses, ${\bf R}$ runs over basis atoms in the unit cell and $\alpha $
runs over directions $x$,$y$,$z$; $\delta _{{\bf R}\alpha }^{+}V$ denotes
self--consistent change in the potential associated with the ${\bf q}$--wave
displacements of atoms ${\bf R}$ along $\alpha $ axe. In practical
calculations we have also added so--called incomplete basis set corrections
to the matrix elements (\ref{f3}) according to the method developed in Ref. %
\onlinecite{EPI}

The coupling strength $\lambda _{{\bf q}\nu }$ for the electrons with the
phonon of wave vector ${\bf q}$ and branch $\nu $ is given by the following
integral 
\begin{equation}
\lambda _{{\bf q}\nu }=\frac 2{N_s(0)}\sum_{{\bf k}jj^{\prime }}\delta (E_{%
{\bf k}j})\delta (E_{{\bf k}+{\bf q}j^{\prime }})|g_{{\bf k}+{\bf q}%
j^{\prime }{\bf k}j}^{{\bf q}\nu }|^2  \label{f2}
\end{equation}
where $N_s(0)$ is the density of states at $E_F$=0 per cell and {\em per one
spin}. Indexes $j$ and $j^{\prime }$ numerate the bands (not spins) and spin
degenerate case is assumed throughout the paper. The total coupling constant 
$\lambda $ results by summing $\lambda _{{\bf q}\nu }$ over $\nu $ and by
averaging over BZ. Two delta functions in (\ref{f2}) impose integration over
the space curve resulting from the crossing of two Fermi surfaces separated
by ${\bf q}$. For this integral we have used as many as 816 ${\bf k}$%
--points per irreducible BZ.

The calculated values of $\lambda _{{\bf q}\nu }$ at the symmetry directions
of the BZ are indicated in Fig. \ref{phn} along with the calculated phonon
dispersions. (We emphasise only the values larger than 0.01.) On top of the
figure shown are the values of $\lambda _{{\bf q}}$ which are summed over
all branches for given ${\bf q}$. It is seen that the electron--phonon
coupling is large for the high--frequency bond--stretching longitudinal
branch. This result is expectable from band structure calculations\cite
{MatHam} since bond--stretching and especially breathing vibrations produce
modest changes in the energy bands near $E_F$. Coupling is strongly enhanced
near the points $M$ and $R$, where it reaches the values $\sim $0.3. Here,
the mode corresponds to two-- or three--dimensional breathing. The value of $%
\lambda _b$=0.3 for the breathing mode at R is in accord with the previous
frozen--phonon calculation\cite{Liecht1}. The authors of Ref. %
\onlinecite{Kunc}, on the other hand, give much lower value for $\lambda _b$
equal to 0.04. For other bond--stretching vibrations, we find $\lambda _{%
{\bf q}\nu }$ of the order 0.1.

From Fig. \ref{phn} we conclude that the electron--phonon coupling is not
small for the bond--bending oxygen modes. It is seen that $\lambda $ is
enhanced for the wave vectors near $R$ and also along $\Gamma X$ direction.
In the latter case, the vibrations correspond to the ferroelectric mode and
the value of $\lambda _{{\bf q}\nu }$ as high as 0.43 is found for the ${\bf %
q}$--point (1/3,0,0)$\pi /a$. Strongly anharmonic tilting modes, on the
other hand, do not exhibit noticeable electron--phonon coupling in the
linear order with respect to the displacements. Exactly at the point $R$,
these modes have electron--phonon matrix element equal to zero by symmetry,
and, therefore, small values of $\omega $ do not bring any effect on
enhancing the coupling. We refer to the following section on our evaluated
anharmonicity corrections.

On the basis of our evaluated phonon dispersion and ${\bf q}$--dependent
electron--phonon interaction we calculate the Eliashberg spectral function $%
\alpha ^2F(\omega )$. This is plotted in Fig. \ref{dos}(b) by full lines.
There, we also show by symbols two $\alpha ^2F(\omega )$ which were deduced
from the tunnelling measurements\cite{TUN1}. Comparing the experiment with
our calculations, it is first seen that the intensities of high--energy
peaks are approximately the same which means that we reproduce the coupling
for these phonons sufficiently accurate. It is also seen both from the
theory and the experiment that while in the phonon density of states [Fig. 
\ref{dos}(a)] mainly the TO\ phonons contribute to the high--energy
structure, for the $\alpha ^2F(\omega )$ these are the LO phonon modes. Our
tendency to overestimate the phonon frequencies at high energies is again
clearly distinguishable.

The most prominent feature seen from our calculated $\alpha ^2F(\omega )$ is
the absence of any structure for the low--frequency interval below 40 meV.
This strongly contradicts with the experimental $\alpha ^2F(\omega )$ which
exhibits two intensive peaks in this region centred at 15 and 30 meV.
According to our analysis of partial $F(\omega )$,\ the origin of the first
peak could be due to low energy Ba(K) and Bi vibrations together with the
tilting modes, and the second peak can result from the bond--bending oxygen
modes. It is not clear however why the calculation seriously underestimates
the electron--phonon coupling for these phonons, while it correctly
describes the coupling for the bond--stretching modes. From the band
structure arguments\cite{MatHam} one can expect that only bond--stretching
modes will have large interaction with electrons. The bond--bending modes
cannot produce any significant changes in the bands near $E_F$ since $%
sp(\sigma )$ interaction for these kind of distortions is not changed in
linear order. The same is true for the tilting modes. Since there is no
partial weight of the Ba(K) orbitals at $E_F$ we also do not expect strong
electron--phonon coupling for the low--frequency interval.

Another possible explanation for the observed peaks is due to the
contributions connected with anharmonicity. For the anharmonic phonons, not
only one--phonon virtual transitions must be seen in the $\alpha ^2F(\omega
) $, but also higher order virtual states. The detailed discussion on this
subject will be presented in the next section, here we only give the value
0.04 as our final answer for the anharmonic contribution to $\lambda $
resulting from the tilting motions. While not negligible, this value alone
again does not explain the intensity of the experimental $\alpha ^2F(\omega
) $ at low--energies.

Our calculated total value of $\lambda $ resulting from the harmonic phonons
and linear electron--phonon coupling is found to be 0.29. {\em This is too
small to account for the superconductivity} {\em at 30K in the compound Ba}$%
_{0.6}${\em K}$_{0.4}${\em BiO}$_3$. In fact, only the high--energy phonons
contribute to our coupling. As a result, our estimated value of $\omega
_{log}$ as high as 550K is found. Using McMillan's $T_c$ expression\cite{McM}
with $\mu ^{*}$=0 we however find the critical temperature with our set of
parameters as low as 4.5K. One can try to estimate the error in our $\lambda
\,\,$value due to the overestimation of the frequencies for the
bond--stretching modes. Using the expression \cite{McM}: $\lambda \simeq
NI^2/M\,\bar \omega ^2$, where $NI^2$ is the electronic prefactor and $M\bar 
\omega ^2$ is an average force constant, one sees that $\lambda $ should
increase with lowering $\bar \omega ^2.$ Our 20\% error in $\bar \omega $
results in 30\% error in $\bar \omega ^2$, and this can lead to the actual $%
\lambda $ values which are 30\% higher than we calculate. However, $\lambda
\sim $0.4 is also not sufficient to explain the value of $T_c$.

Our calculated electron--phonon contribution to the transport properties is
a final subject of this section. The quantities responsible for the
electronic transport are easily deduced from the linear--response
calculations \cite{EPI}. By inserting the electron velocity factor $(v_{{\bf %
k}j}-v_{{\bf k}+{\bf q}j^{\prime }})^2$ to the expression (\ref{f2}), we
calculate transport constant $\lambda _{tr}$. Its ${\bf q}$--dependence is
shown at the top of Fig. \ref{phn} along with the ${\bf q}$--dependence of
the electron--phonon $\lambda $. We see that both functions exhibit very
similar behaviour in the BZ. The transport spectral function $\alpha
_{tr}^2F(\omega )$ is the central quantity needed for evaluating
temperature--dependent electrical and thermal resistivity as low--order
variational solutions of the Boltzmann equation \cite{Allen}. Our calculated $%
\alpha _{tr}^2F(\omega )$ is shown in Fig. \ref{dos}(c). It is seen that
this function behaves closely to the superconducting $\alpha ^2F(\omega )$
which is usually the case in metals\cite{EPI}. The total average $\lambda
_{tr}$ is found to be 0.32. Based on these data, we evaluate
electron--phonon limited electrical resistivity $\rho $ to be 14.3 $\mu
\Omega \times cm$ at $T$=273K. This is at least one order of magnitude lower
than the values of $\rho $ reported in the literature\cite{RES}. It is clear
that the source for this discrepancy is the same as in our describing
superconducting properties. It is unlikely that the strong electronic
correlations are presented in these materials because the parent compounds
are diamagnetic (not antiferromagnetic) insulators. Therefore, it is
unlikely that other (than electron--phonon) scattering mechanisms take
place, as spin fluctuations, for example, in HTSC cuprates. Taking into
account anharmonic phonons, polarons, or bipolarons may be decisive for
describing the superconductivity and transport phenomena here. Our basic
conclusion is that {\em the conventional ideas on the electron--phonon
mechanism are not operative in the bismuthates}.

\begin{center}
{\bf III. TILTING\ AND\ ANHARMONIC\ }$\lambda .$
\end{center}

Towards further understanding of superconductivity in the bismuthates, we
try to evaluate anharmonicity corrections in the electron--phonon
interaction. As we have mentioned in the introduction, several experiments
and numerous frozen--phonon calculations point on possible importance of
these effects in the theory of HTSC cuprates\cite
{Pickett,Krak,Hardy,Schuttler,Cohen}. In the bismuthates, the first
candidate to study anharmonicity is the tilting mode: Our linear--response
calculations give nearly zero--frequency vibrations for this mode at the $R$%
--point of cubic BZ. Our own (and previous\cite{Kunc}) frozen--phonon
calculations predict here a double--potential--well behaviour with very
shallow energy minimum at 5$^{\circ }$ as illustrated in Fig. \ref{tilt}.
Despite the reported average structure for the superconducting phase is
cubic \cite{STRUC2}, some experiments \cite{Braden} discuss the existence of
long--range superstructure characterised by the rotations of the octahedra.
Recent XAFS measurements \cite{Jacoby} report on the presence of locally
disordered rotations.

Unfortunately, though formulated \cite{Maks}, the problem of the influence
of anharmonicity to superconductivity is not tractable numerically in a full
volume. Our simplified treatment is based on the expression introduced by
Hui and Allen \cite{Hui} which generalises zero--temperature electron-phonon
coupling to the anharmonic case by including matrix elements over all phonon
excited states. The phonon states $|n\rangle $ and their energies $\epsilon
_n$ are obtained by solving the Schr\"odinger equation for an oscillatory
mode characterised by ${\bf q}$ and $\nu $. (We will not label the states $%
|n\rangle ,\epsilon _n$ with (${\bf q}\nu )$ for simplicity). For harmonic
potential wells $\epsilon _n-\epsilon _0$ is just $n\omega _{{\bf q}\nu }$
(in atomic units), where $\omega _{{\bf q}\nu }$ is the phonon frequency.
This leads only to the one--phonon virtual states ($n$=1) which are involved
in the matrix elements of the electron--phonon interaction. For anharmonic
potential wells the spectrum is generally different from the set of
equidistant levels. One example is a double well of the tilting mode which
is shown in Fig. \ref{tilt} The real spectrum $\epsilon _n$ obtained as the
solution of the Schr\"odinger equation is plotted in Fig. \ref{tilt} by
horizontal lines. Therefore, our first purpose is to examine what effect in
our $\lambda $ would bring the proper treatment of all virtual phonon
states. The second problem is connected with the modification of the
electron--phonon matrix elements due to higher--order terms in the expansion
of the change in the ground state potential with respect to the atomic
displacements. Since we wish to examine these effects only for the tilting
mode at the point $R$ of the cubic BZ, we make an essential approximation by
assuming that the tilting mode is not coupled to the other modes of either
this wave vector or other wave vectors which is generally not true when
anharmonic terms are included into the lattice dynamical problem. We assume
that the polarisation vectors for this mode are known and are given by
exactly the out--of--phase rotations of the nearest octahedra along the
(1,1,0) axe. Any processes of phonon--phonon interactions will be neglected
in this treatment. We also neglect by the finite--temperature effects using
the arguments given in Ref. \onlinecite{Cohen} In the double--well problem
one expects that the modifications due to frequency and
electron--phonon--matrix--elements renormalisation are quite dramatic \cite
{Hardy,Cohen} and bring the largest effect in the values of $\lambda $.

\end{multicols}

We start from a general zero--temperature expression for the
electron--phonon coupling in anharmonic case, which can be written as
follows \cite{Hui}

\begin{equation}
\lambda _{{\bf q}\nu }=\frac 1{N_s(0)}\sum_{{\bf k}^{\prime }j^{\prime }{\bf %
k}j}\sum_n(f_{{\bf k}j}-f_{{\bf k}^{\prime }j^{\prime }})\delta (E_{{\bf k}%
j}-E_{{\bf k}^{\prime }j^{\prime }}+\omega _n)\left| G_{{\bf k}^{\prime
}j^{\prime }{\bf k}j}^{[n]}\right| ^2/[\omega _n]^2,  \label{a1}
\end{equation}
where $\omega _n=\epsilon _n-\epsilon _0$. For the moment we will not assume
as in Ref. \onlinecite{Hui} that $\omega _n$ are small at the electron
energy scale. (The latter reduces the integral with the Fermi step functions 
$f_{{\bf k}j}-f_{{\bf k}^{\prime }j^{\prime }}$ to the integral with the
delta functions as given by Eq. \ref{f2}.) We introduce a generalised matrix
element $G_{{\bf k}^{\prime }j^{\prime }{\bf k}j}^{[n]}\,$ for the virtual
transition to the $n$'s phonon state: 
\begin{equation}
G_{{\bf k}^{\prime }j^{\prime }{\bf k}j}^{[n]}=\langle n|\Delta g_{{\bf k}%
^{\prime }j^{\prime }{\bf k}j}|0\rangle .  \label{as1}
\end{equation}
The electron--phonon matrix element $\Delta g_{{\bf k}^{\prime }j^{\prime }%
{\bf k}j}$ in (\ref{as1}) involves the transitions between the states $|{\bf %
k}j\rangle $ and $|{\bf k}^{\prime }j^{\prime }\rangle $ near the Fermi
surface

\begin{equation}
\Delta g_{{\bf k}^{\prime }j^{\prime }{\bf k}j}=\langle {\bf k}^{\prime
}j^{\prime }|\Delta V|{\bf k}j\rangle ,  \label{a2}
\end{equation}
where $\Delta V$ is the total change in the ground state potential induced
by the lattice distortion associated with the mode ${\bf q}\nu $. We have
especially included $\Delta $ in the notation $\Delta g_{{\bf k}^{\prime
}j^{\prime }{\bf k}j}$ since $\Delta V$ is not a derivative of the potential
with respect to the displacements but it is the difference between the
self--consistent potential $V({\bf r},\{{\bf t}_{{\bf R}}+\Delta {\bf t}_{%
{\bf R}}\})$ for the distorted crystal and the potential $V({\bf r},\{{\bf t}%
_{{\bf R}}\})$ for the undistorted crystal. The atomic positions at the
equilibrium are given by the vectors ${\bf t}_{{\bf R}}={\bf t+R}$, where $%
{\bf t}$ denote the translations and ${\bf R}$ are the basis vectors. The
displacements associated with the mode ${\bf q}\nu $ are described by the
vector field $\Delta {\bf t}_{{\bf R}}$. Therefore $\Delta V$ is
proportional to $\Delta {\bf t}_{{\bf R}}$ and so does $\Delta g_{{\bf k}%
^{\prime }j^{\prime }{\bf k}j}$. By introducing complex (infinitesimal)
polarisation vectors $\delta {\bf Q}_R$ of the mode \cite{tilt} ${\bf q}\nu $%
, the displacements in any atomic cell ${\bf t}$ can be found using the
formula:

\begin{equation}
\Delta t_{{\bf R}\alpha }=\delta Q_{{\bf R}\alpha }e^{i{\bf qt}}+c.c.
\label{a3}
\end{equation}
where $\alpha $ runs over directions $x,y,z$, and $c.c.$ stands for the
complex conjugate. (The quantities $\Delta V$, $\Delta {\bf t}_{{\bf R}}$,
and $\delta {\bf Q}_R$ should, in principle, be labelled with ${\bf q}\nu $
but we omit this for simplicity.)

The phonon states $|n\rangle $ are the functions of the displacements $%
\Delta t_{{\bf R}\alpha }$ or $\delta Q_{{\bf R}\alpha }$. In order to
compute the matrix element $\langle n|\Delta g_{{\bf k}^{\prime }j^{\prime }%
{\bf k}j}|0\rangle \,$ over the phonon states we should expand $\Delta V$
with respect to the displacements. Keeping the terms up to second order,
this expansion reads as

\begin{eqnarray}
\ &&\Delta V=\sum_{{\bf R}\alpha }\delta Q_{{\bf R}\alpha }\sum_{{\bf t}}e^{i%
{\bf qt}}\frac{\delta V}{\delta t_{{\bf R}\alpha }}+  \nonumber \\
&&\ \ \ \ \ +\frac 12\sum_{{\bf RR}^{\prime }\alpha \alpha ^{\prime }}\delta
Q_{{\bf R}\alpha }\delta Q_{{\bf R}^{\prime }\alpha ^{\prime }}\sum_{{\bf tt}%
^{\prime }}e^{i{\bf q(t+t}^{\prime })}\frac{\delta ^{(2)}V}{\delta t_{{\bf R}%
\alpha }\delta t_{{\bf R}^{\prime }\alpha ^{\prime }}^{\prime }}+  \nonumber
\\
&&\ \ \ \ \ +\frac 12\sum_{{\bf RR}^{\prime }\alpha \alpha ^{\prime }}\delta
Q_{{\bf R}\alpha }(\delta Q_{{\bf R}^{\prime }\alpha ^{\prime }})^{*}\sum_{%
{\bf tt}^{\prime }}e^{i{\bf q(t-t}^{\prime })}\frac{\delta ^{(2)}V}{\delta
t_{{\bf R}\alpha }\delta t_{{\bf R}^{\prime }\alpha ^{\prime }}^{\prime }}%
+c.c.  \label{a4}
\end{eqnarray}
Here, $\delta V/\delta t_{{\bf R}\alpha }$ is associated with the
first--order derivative of the potential when a single nucleus centred at $%
{\bf t}+{\bf R}$ experiences an infinitesimal displacement along $\alpha $%
--th direction, and $\delta ^{(2)}V/\delta t_{{\bf R}\alpha }\delta t_{{\bf R%
}^{\prime }\alpha ^{\prime }}^{\prime }$ stands for the second--order
derivative. [Notation $t_{{\bf R}\alpha }$ is shorthand for $(t+R)_\alpha .$%
] Obviously, both these response functions have no dependence on the mode $%
{\bf q}\nu .$ If $V({\bf r})$ has a periodicity of the original lattice, the
change $\delta V/\delta t_{{\bf R}\alpha }$ is a function of general type.
One expects that $\delta V/\delta t_{{\bf R}\alpha }$ is only not zero in
the vicinity of the displaced atom and it goes to zero when ${\bf r}$
departs from the site ${\bf t}+{\bf R}$. However, because of the
translational invariance of the original crystal, considering the response
at the point ${\bf r}$ due to the movement of atom in ${\bf t}+{\bf R}$ must
be equivalent to considering the response at the point ${\bf r}-{\bf t}$ due
to the movement of atom at ${\bf R}$ (when ${\bf t}$=0). Therefore we can
write that $\delta V({\bf r})/\delta t_{{\bf R}\alpha }=\delta V({\bf r}-%
{\bf t})/\delta R_\alpha .$ We now introduce the lattice sum

\begin{equation}
\delta _{{\bf R}\alpha }^{+}V=\sum_{{\bf t}}e^{i{\bf qt}}\frac{\delta V}{%
\delta t_{{\bf R}\alpha }}  \label{a5}
\end{equation}
which represent a variation of the potential per unit displacement induced
by to the movements of atoms ${\bf R}$ along $\alpha $--th axe by
infinitesimal amount $\delta t_{{\bf R}\alpha }$ proportional to $exp(i{\bf %
qt})$ in every cell ${\bf t}.$ It is easy to prove that the expression (\ref
{a5}) translates like a Bloch wave with wave vector ${\bf q}$ in the
original lattice, i.e. $\delta ^{+}V({\bf r}+{\bf R})=e^{i{\bf qR}}\delta
^{+}V({\bf r})$. (We will sometimes omit indexes ${\bf R}\alpha )$ Notation $%
\delta ^{+}V$ refers to the travelling wave of vector $+{\bf q}$ while
complex conjugated quantity $\delta ^{-}V$ would refer to the travelling
wave of vector $-{\bf q}$.

One can analogously define lattice sums associated with the second--order
changes of the potential. These enter the second and third contributions in (%
\ref{a5}). Consider, for example, the lattice sum associated with the second
contribution: 
\begin{equation}
\delta _{{\bf R}\alpha }^{+}\delta _{{\bf R}^{\prime }\alpha ^{\prime
}}^{+}V=\sum_{{\bf tt}^{\prime }}e^{i{\bf q(t}+{\bf t}^{\prime })}\frac{%
\delta ^{(2)}V}{\delta t_{{\bf R}\alpha }\delta t_{{\bf R}^{\prime }\alpha
^{\prime }}^{\prime }}.  \label{a6}
\end{equation}
This expression translates like a Bloch wave of vector $2{\bf q}$ because 
\begin{eqnarray}
&&\sum_{{\bf tt}^{\prime }}e^{i{\bf q(t}+{\bf t}^{\prime })}\frac{\delta
^{(2)}V({\bf r}+{\bf t}^{\prime \prime })}{\delta t_{{\bf R}\alpha }\delta
t_{{\bf R}^{\prime }\alpha ^{\prime }}^{\prime }}=  \nonumber \\
&&\sum_{{\bf tt}^{\prime }}e^{i{\bf q(t}+{\bf t}^{\prime })}\frac{\delta
^{(2)}V({\bf r})}{\delta (t-t^{\prime \prime })_{{\bf R}\alpha }\delta
(t^{\prime }-t^{\prime \prime })_{{\bf R}^{\prime }\alpha ^{\prime }}}= 
\nonumber \\
&&e^{2i{\bf q}t^{\prime \prime }}\sum_{{\bf tt}^{\prime }}e^{i{\bf q(t}+{\bf %
t}^{\prime })}\frac{\delta ^{(2)}V({\bf r})}{\delta t_{{\bf R}\alpha }\delta
t_{{\bf R}^{\prime }\alpha ^{\prime }}^{\prime }}.  \label{a7}
\end{eqnarray}
Analogously, the lattice sum associated with the third contribution in (\ref
{a5}) can be denoted as $\delta ^{+}\delta ^{-}V$. It represents a
travelling wave of wave vector ${\bf 0}$, i.e. it is periodical at original
lattice.

Using the notations (\ref{a5}), and (\ref{a6}), the change in the potential $%
\Delta V$ given by the formula (\ref{a4}) now has the form 
\begin{eqnarray}
\Delta V &=&\sum_{{\bf R}\alpha }\delta Q_{{\bf R}\alpha }\times \delta _{%
{\bf R}\alpha }^{+}V+  \nonumber \\
&&\ \ \ +\frac 12\sum_{{\bf RR}^{\prime }\alpha \alpha ^{\prime }}\delta Q_{%
{\bf R}\alpha }\delta Q_{{\bf R}^{\prime }\alpha ^{\prime }}\times \delta _{%
{\bf R}\alpha }^{+}\delta _{{\bf R}^{\prime }\alpha ^{\prime }}^{+}V+ 
\nonumber \\
&&\ \ \ +\frac 12\sum_{{\bf RR}^{\prime }\alpha \alpha ^{\prime }}\delta Q_{%
{\bf R}\alpha }(\delta Q_{{\bf R}^{\prime }\alpha ^{\prime }})^{*}\times
\delta _{{\bf R}\alpha }^{+}\delta _{{\bf R}^{\prime }\alpha ^{\prime
}}^{-}V+c.c.  \label{a8}
\end{eqnarray}
It is clear that when this expansion is used in the matrix element (\ref{a2}%
), certain selection rule will occur for the wave vectors ${\bf k}^{\prime }$
and ${\bf k}$. Namely, the matrix element $\langle {\bf k}^{\prime
}j^{\prime }|\delta ^{+}V|{\bf k}j\rangle $ is equal to zero unless ${\bf k}%
^{\prime }={\bf k}+{\bf q}$, $\langle {\bf k}^{\prime }j^{\prime }|\delta
^{+}\delta ^{+}V|{\bf k}j\rangle $=0 unless ${\bf k}^{\prime }={\bf k}+2{\bf %
q}$, and $\langle {\bf k}^{\prime }j^{\prime }|\delta ^{+}\delta ^{-}V|{\bf k%
}j\rangle $=0 unless ${\bf k}^{\prime }={\bf k}$.

Let us now introduce the electron--phonon matrix element associated with the
first--order change in the potential

\begin{equation}
G_{{\bf k}^{\prime }j^{\prime }{\bf k}j}^{[n]\{1\}}=\delta _{{\bf k}^{\prime
}{\bf k}+{\bf q}}\left\langle {\bf k}+{\bf q}j^{\prime }\left| \sum_{{\bf R}%
\alpha }\langle n|\delta Q_{{\bf R}\alpha }|0\rangle \times \delta _{{\bf R}%
\alpha }^{+}V\right| {\bf k}j\right\rangle .  \label{a9}
\end{equation}
The electron--phonon matrix elements associated with the second--order
changes in the potential have two forms according to the second and third
contributions in (\ref{a8}) 
\begin{equation}
G_{{\bf k}^{\prime }j^{\prime }{\bf k}j}^{[n]\{2\}}=\frac 12\delta _{{\bf k}%
^{\prime }{\bf k}+2{\bf q}}\left\langle {\bf k}+2{\bf q}j^{\prime }\left|
\sum_{{\bf RR}^{\prime }\alpha \alpha ^{\prime }}\langle n|\delta Q_{{\bf R}%
\alpha }\delta Q_{{\bf R}^{\prime }\alpha ^{\prime }}|0\rangle \times \delta
_{{\bf R}\alpha }^{+}\delta _{{\bf R}^{\prime }\alpha ^{\prime
}}^{+}V\right| {\bf k}j\right\rangle ,  \label{a10}
\end{equation}
\begin{equation}
G_{{\bf k}^{\prime }j^{\prime }{\bf k}j}^{[n]\{2^{\prime }\}}=\frac 12\delta
_{{\bf k}^{\prime }{\bf k}}\left\langle {\bf k}j^{\prime }\left| \sum_{{\bf %
RR}^{\prime }\alpha \alpha ^{\prime }}\langle n|\delta Q_{{\bf R}\alpha
}(\delta Q_{{\bf R}^{\prime }\alpha ^{\prime }})^{*}|0\rangle \times \delta
_{{\bf R}\alpha }^{+}\delta _{{\bf R}^{\prime }\alpha ^{\prime
}}^{-}V\right| {\bf k}j\right\rangle .  \label{a11}
\end{equation}
Then, the expression (\ref{a1}) for $\lambda _{{\bf q}\nu }$ splits into
three contributions

\begin{equation}
\lambda _{{\bf q}\nu }=\lambda _{{\bf q}\nu }^{\{1\}}+\lambda _{{\bf q}\nu
}^{\{2\}}+\lambda _{{\bf q}\nu }^{\{2^{\prime }\}}  \label{a12}
\end{equation}
associated with one electron--phonon matrix element from the first order,
Eq. (\ref{a9}), and two matrix elements from the second--order, Eqs. (\ref
{a10}), and (\ref{a11}), i.e.

\begin{eqnarray}
\lambda _{{\bf q}\nu }^{\{1\}} &=&\frac 2{N_s(0)}\sum_{{\bf k}jj^{\prime
}}\sum_n(f_{{\bf k}j}-f_{{\bf k}+{\bf q}j^{\prime }})\delta (E_{{\bf k}j}-E_{%
{\bf k}+{\bf q}j^{\prime }}+\omega _n)\left| G_{{\bf k}+{\bf q}j^{\prime }%
{\bf k}j}^{[n]\{1\}}\right| ^2/[\omega _n]^2,  \label{a13} \\
\lambda _{{\bf q}\nu }^{\{2\}} &=&\frac 2{N_s(0)}\sum_{{\bf k}jj^{\prime
}}\sum_n(f_{{\bf k}j}-f_{{\bf k}+2{\bf q}j^{\prime }})\delta (E_{{\bf k}%
j}-E_{{\bf k}+2{\bf q}j^{\prime }}+\omega _n)\left| G_{{\bf k}+2{\bf q}%
j^{\prime }{\bf k}j}^{[n]\{2\}}\right| ^2/[\omega _n]^2,  \label{a14} \\
\lambda _{{\bf q}\nu }^{\{2^{\prime }\}} &=&\frac 2{N_s(0)}\sum_{{\bf k}%
jj^{\prime }}\sum_n(f_{{\bf k}j}-f_{{\bf k}j^{\prime }})\delta (E_{{\bf k}%
j}-E_{{\bf k}j^{\prime }}+\omega _n)\left| G_{{\bf k}j^{\prime }{\bf k}%
j}^{[n]\{2^{\prime }\}}\right| ^2/[\omega _n]^2.  \label{a15}
\end{eqnarray}
The double sums over ${\bf k}$ and ${\bf k}^{\prime }$ appeared in (\ref{a1}%
) are now reduced to the single sums over ${\bf k}$ according to the
selection rules in the matrix elements (\ref{a9}), (\ref{a10}), and (\ref
{a11}). Taking into account matrix elements coming from the complex
conjugated quantities $\delta ^{-}V$, $\delta ^{-}\delta ^{-}V$, and $\delta
^{-}\delta ^{+}V$ gives contributions with wave vectors ${\bf k}-{\bf q}$, $%
{\bf k}-2{\bf q,\,}$and ${\bf k.}$ However making substitutions ${\bf k}-%
{\bf q}\rightarrow {\bf k}\,$ and ${\bf k}-2{\bf q}\rightarrow {\bf k}\,$ in
the ${\bf k}$--space integrals of (\ref{a1}) results in extra factor of 2
which appears in (\ref{a13}), (\ref{a14}), and (\ref{a15}) compared to (\ref
{a1}).

We now discuss the derived expressions. First, it is instructive to see how
Eq.\ref{a13} for $\lambda _{{\bf q}\nu }^{\{1\}}$ goes to the standard
formula (\ref{f2})$.$ For the case of harmonic oscillator we can use the
properties of Hermite polynomials and prove that the matrix element over the
phonon states appeared in (\ref{a9}) is reduced to

\begin{equation}
\langle n|\delta Q_{{\bf R}\alpha }|0\rangle =\delta _{n1}\frac{Q_{{\bf R}%
\alpha }}{\sqrt{2M_{{\bf R}}\omega _{{\bf q}\nu }}},  \label{as9}
\end{equation}
where $Q_{{\bf R}\alpha }$ denote ortho normalised polarisation vectors of
the mode ${\bf q}\nu \,$ in contrast to $\delta Q_{{\bf R}\alpha }$ which
are related to the actual nuclei displacements \cite{tilt}. The matrix
element (\ref{as9}) is only not zero for the transitions including
one--phonon virtual state. Substituting (\ref{as9}) into (\ref{a9}) gives
exactly the formula (\ref{f3}). By placing (\ref{a9}) to (\ref{a13}) we see
that only $n$=1 term in the sum over $n$ survives, and the energy difference 
$\omega _1$ is equal to $\omega _{{\bf q}\nu }$. The standard formula (\ref
{f2}) is then recovered \cite{comm1} by replacing ($f_{{\bf k}j}-f_{{\bf k}%
^{\prime }j^{\prime }})\delta (E_{{\bf k}j}-E_{{\bf k}+{\bf q}j^{\prime
}}+\omega _{{\bf q}\nu })$ to $\omega _{{\bf q}\nu }\delta (E_{{\bf k}%
j})\delta (E_{{\bf k}+{\bf q}j^{\prime }})$ which is valid for small $\omega
_{{\bf q}\nu }$.

Second, it should be noted that the number of matrix elements necessary to
reach the convergency in the sum over $n$ is actually not large. From the
numerical estimates of the oscillator strengths $f_n$ for the transitions
from the phonon ground state $|0\rangle $ to excited states $|n\rangle ,$
the authors of Ref. \onlinecite{Hui} concluded that the f--sum rule (sum
over all oscillator strengths gives unity) is closely satisfied by taking
into account only single term $f_{1.}$ Even for infinite rectangular well
which represents an extreme case for anharmonicity, $n$=5 is sufficient\cite
{Hui}. Our own numerical experiments with the double--potential well of the
form shown in Fig. \ref{tilt} confirms this conclusion both for the dipole
matrix elements and for quadrupole ones. This implies that the phonon
excitation energies $\omega _n=\epsilon _n-\epsilon _0$ appeared in the
expressions (\ref{a13}), (\ref{a14}), and (\ref{a15}) are still too small at
the electronic energy scale and, therefore, all integrals over ${\bf k}$ are
reduced to the integrals over the space curve resulting from the crossing of
two Fermi surfaces separated by ${\bf q}$. It then follows that {\em the
expression for} $\lambda _{{\bf q}\nu }^{\{2^{\prime }\}}$ {\em is} {\em %
always equal to zero} unless there are electronic inter band transitions with
zero momentum transfer at the phonon energies. The latter standardly is not
assumed. We therefore left with the following formulae for $\lambda _{{\bf q}%
\nu }^{\{1\}}$ and $\lambda _{{\bf q}\nu }^{\{2\}}$:

\begin{eqnarray}
\lambda _{{\bf q}\nu }^{\{1\}} &=&\frac 2{N_s(0)}\sum_{{\bf k}jj^{\prime
}}\delta (E_{{\bf k}j})\delta (E_{{\bf k}+{\bf q}j^{\prime }})\sum_n\left|
G_{{\bf k}+{\bf q}j^{\prime }{\bf k}j}^{[n]\{1\}}\right| ^2/\omega _n,
\label{a16} \\
\lambda _{{\bf q}\nu }^{\{2\}} &=&\frac 2{N_s(0)}\sum_{{\bf k}jj^{\prime
}}\delta (E_{{\bf k}j})\delta (E_{{\bf k}+2{\bf q}j^{\prime }})\sum_n\left|
G_{{\bf k}+2{\bf q}j^{\prime }{\bf k}j}^{[n]\{2\}}\right| ^2/\omega _n.
\label{a17}
\end{eqnarray}

Another comment concerns the expressions (\ref{a14}) and (\ref{a17}) for $%
\lambda _{{\bf q}\nu }^{\{2\}}$ which involves electronic int{\em ra}band
transitions with momentum transfer equal to $2{\bf q}$. If wave vector ${\bf %
q}$ is such zone--boundary vector that $2{\bf q}={\bf G}$, where ${\bf G}$
is one of the reciprocal lattice vectors, then it is obvious that $\lambda _{%
{\bf q}\nu }^{\{2\}}$=0 according to the arguments given above. The
situation here is analogous to the discussion of contributions to $\lambda
\, $ from the $\Gamma $ optical phonons (see, for example Ref. %
\onlinecite{Rodriguez,Ole}). It is well known that exactly ${\bf q}$=0--phonons do not
couple to the electrons since the energy cannot be conserved. Therefore,
small but finite vectors $q\sim \omega _{q\nu }/v_F$, where $v_F$ is the
Fermi velocity, must be considered.

\begin{multicols}{2}

We now discuss implications of the theory given above for the tilting mode
in Ba$_{0.6}$K$_{0.4}$BiO$_3$. Exactly for the ${\bf q}$--point $R$=(1,1,1)$%
\pi /a$ the contribution to $\lambda $ connected with the first--order
changes in the potential, $\lambda _{{\bf q}\nu }^{\{1\}}$, vanishes due to
symmetry of the matrix elements. Also the contribution connected with the
second--order changes, $\lambda _{{\bf q}\nu }^{\{2\}}$, is zero since
vector $2{\bf q\,}$ is equivalent to the $\Gamma $ point. The contribution $%
\lambda _{{\bf q}\nu }^{\{2^{\prime }\}}$ is always zero since there is only
one band crossing $E_F$ and there is no interband transitions near $E_F$. In
the higher orders, there are contributions to $\lambda $ connected with the
odd--order changes in the potential and with the even--order changes. In the
odd orders, changes in the potential would represent travelling waves of
vectors ${\bf q},3{\bf q},5{\bf q},$ etc., and the contributions to $\lambda 
$ would be equal to zero due to symmetry of the matrix elements. In the even
orders, changes in the potential would represent travelling waves of vectors 
${\bf 0},2{\bf q},4{\bf q},$ etc., which are all equivalent to the $\Gamma $
point when ${\bf q}$=(1,1,1)$\pi /a$, and corresponding contributions to $%
\lambda $ also vanish. Therefore, we conclude that {\em pure rotational mode
at }$R${\em \ does not couple to the electrons in any order}.

In order to obtain an estimate of the effect, we must, in principle, step
out from the point $R$. Unfortunately, for general ${\bf q}$ we cannot
perform frozen--phonon calculations of anharmonic coefficients, and the
problem looses its numerical tractability. Another approach is to discuss
quantities resulting from the integration of the expressions (\ref{a16}),
and (\ref{a17}) over ${\bf q}$ . In this way, we should assume that the
matrix elements $G_{{\bf k}+{\bf q}j^{\prime }{\bf k}j}^{[n]\{1\}}$ and $G_{%
{\bf k}+2{\bf q}j^{\prime }{\bf k}j}^{[n]\{2\}}$ have no ${\bf q}$%
--dependence (which is generally not true but the order of magnitude of the
effect will be certainly captured) and can be approximated by the values at
the point $R$. Therefore, the contribution resulting from the integration of 
$\lambda _{{\bf q}\nu }^{\{1\}}$ disappears due to symmetry of $G_{{\bf k}+%
{\bf q}j^{\prime }{\bf k}j}^{[n]\{1\}}$ at $R$, and integration of $\lambda
_{{\bf q}\nu }^{\{2\}}$ over ${\bf q}$ gives:

\begin{equation}
\lambda _\nu ^{\{2\}}=2\sum_{{\bf k}j}\delta (E_{{\bf k}j})\sum_n\left| G_{%
{\bf k}j{\bf k}j}^{[n]\{2\}}\right| ^2/\omega _n.  \label{a18}
\end{equation}
(Here we have assumed that ${\bf k}+2{\bf q}$ is equivalent to ${\bf k}$.)
The obtained expression is an average of squared deformation potential which
is induced by the tilting of the octahedra in the second order with respect
to the displacements. It has the same meaning as discussed in the literature 
\cite{Rodriguez,Ole} magnitude of the change of the phonon self--energy due to
transitions across the gap in the superconducting state (so called
superconducting $\lambda _s$).

The value of $\lambda _\nu ^{\{2\}}$ can be readily evaluated using the
frozen--phonon method. Let us measure \cite{tilt} the tilting distortions by
tilting angle $t.$ Quadrupole matrix elements between the phonon states are
then given by $\langle n|t^2|0\rangle $ and the electron--phonon matrix
elements are $\langle {\bf k}j|\delta ^{(2)}V/\delta t^2|{\bf k}j\rangle $
where $\delta ^{(2)}V/\delta t^2$ is the second--order derivative of the
potential with respect to the tilting angle taken at $t$=0.  The matrix
elements $\langle {\bf k}j|\delta ^{(2)}V/\delta t^2|{\bf k}j\rangle $ can
be approximated \cite{comm2} by the second--order derivatives of the
one--electron energies $E_{{\bf k}j}$ relative to $E_{F\text{ }.}$ They are
accessible from band structure calculations. Using this, the formula (\ref
{a18}) for $\lambda _\nu ^{\{2\}}$ can be rewritten as follows

\begin{equation}
\lambda _\nu ^{\{2\}}=2\sum_{{\bf k}j}\delta (E_{{\bf k}j})\frac{d^{(2)}(E_{%
{\bf k}j})^2}{dt^2}\times \sum_n\frac{|\langle n|t^2|0\rangle |^2}{\omega _n}
\label{a19}
\end{equation}
where the quadrupole matrix elements and the energy differences $\omega _n$
are obtained by solving the Schr\"odinger equation for the double potential
well shown in Fig. \ref{tilt}. Only even--order matrix elements $\langle
n=2k|t^2|0\rangle $ are allowed by symmetry of our distortion, and
therefore, the transition between nearly degenerate first and second states
shown in Fig. \ref{tilt} (i.e. $\omega _1\rightarrow 0$) does not contribute 
\cite{comm3} to $\lambda _\nu ^{\{2\}}$. 

In order to evaluate (\ref{a19}), we first calculate energy bands $E_{{\bf k}%
j}[t]$ relative to $E_F[t]$ for a set of distortions $t$, and then integrate

\begin{equation}
F(t)=2\sum_{{\bf k}j}\delta \left( E_{{\bf k}j}[0]\right) \times \left( E_{%
{\bf k}j}[t]\right) ^2  \label{a20}
\end{equation}
over the unperturbed Fermi surface. Due to symmetry, the energy bands $E_{%
{\bf k}j}[t]$ have only even terms in the Taylor's expansion over $t$ around $t
$=0. The function $F(t)-F(0)\,$ is fitted to the polynomial starting with $%
at^4.$ Then, our $\lambda _\nu ^{\{2\}}$ is

\begin{equation}
\lambda _\nu ^{\{2\}}=a\sum_{2n}\frac{|\langle 2n|t^2|0\rangle |^2}{\omega
_{2n}}.  \label{a21}
\end{equation}
Using frozen--phonon calculation, we evaluate $a$=2.991 Ry/rad$^4$. Solving
the Schr\"odinger equation for the double well, we also evaluate the sum
over $n$ in (\ref{a21}) equal to 0.0044 rad$^4/$Ry. This gives the total
estimated $\lambda _\nu ^{\{2\}}$=0.013. Since the tilting mode is triple
degenerate, the final result $\lambda ^{\{2\}}=3\lambda _\nu ^{\{2\}}\sim
0.04$ is obtained. It can be compared with the value 0.11 previously
reported by Kunc and Zeyher \cite{Kunc} who used a somewhat analogous
approach. Unfortunately no details were presented in Ref. \onlinecite{Kunc}
and, therefore, it is difficult to determine the main source of our
discrepancies.

Despite several approximations were made in the above derivation, we think
that the values of the order $0.04$ (to 0.11) can be realistic for
anharmonic contribution to $\lambda $ from the tilting modes involving large
ionic excursions. This constitutes about 20\% of the $\lambda $ value 0.29
found by the linear--response method. Unfortunately, our total harmonic plus
anharmonic $\lambda _{tot}$=0.33 is still too small to account for the
superconductivity at 30K in Ba$_{0.6}$K$_{0.4}$BiO$_3$.

In order to have large effect in $\lambda $ due to anharmonicity, one has to
analyse three contributions in Eq. \ref{a19}. The first one is connected
with the changes in the energy bands. The second contribution goes from the
matrix elements $|\langle n|t^2|0\rangle |^2\,\,$which can be large if large
atomic displacements are possible due to the flatness of the potential well
of the oscillator. This is exactly the case for tilting distortions. The
third contribution can go from the energy denominator in case values $\omega
_n$=$\epsilon _n-\epsilon _0\,\,$become sufficiently small \cite{comm3}.

The fact that our anharmonic $\lambda ^{\{2\}}$ is small is due to small
changes in the one--electron energies associated with tilting. This is so
because the tilting distortions do not change the distance between Bi and O
atoms, thus keeping $sp(\sigma )$ nearest--neighbour interaction nearly
constant. Possible large contribution from the energy bands would be
expected for breathing distortions. Unfortunately we have not found strong
anharmonicity for the breathing potential well in doped BaBiO$_3$. It is
however known that in the undoped compound there exist frozen breathing
distortions which correspond to a deep--double--well situation. It could be
possible that our LDA description of breathing distortions is not completely
correct. The following section is devoted to this question.

\begin{center}
{\bf IV. BREATHING\ AND\ LDA}
\end{center}

So far we have discussed the compound Ba$_{0.6}$K$_{0.4}$BiO$_3$ which is a
metal in its normal state. Our calculations show that such ground state
properties as equilibrium structure and lattice dynamics in the adiabatic
approximation are reproduced by the density functional LDA\ method
reasonably well. This suggests that our description of the electron--phonon
interactions based on the LDA energy bands is realistic. We now turn to the
discussion of the undoped parent compound BaBiO$_3.$ It is known for many
years that pure BaBiO$_3$ is a charge--density--wave semiconductor in a
sense that two bismuth atoms exist in the charge disproportionate state Bi$%
^{4\pm q}$. This leads to the modulated Bi--O distances (breathing
distortions) in the cubic perovskite lattice. In addition, there are strong
tilting distortions \cite{STRUC}. The nature of the disproportionate state
is still not very well understood and several explanations involving Fermi
surface nesting \cite{MatHam}, a real space pairing based on strong
electron--phonon interaction \cite{Rice}, and the existence of negative
electronic $U$ due to Bi$^{4+}$ valence skipping \cite{Varma} have been
suggested in the past. It is also unclear whether there is any connection
between the origin of the semiconducting behaviour and the doping induced
superconductivity a the border of metal--insulator transition.

If intra atomic correlations of Bi $6s$ electrons are strong and responsible
for the appearance of negative $U$, one expects that the LDA theory will
fail to describe the charge--disproportionate state. This exactly happens
in HTSC cuprates where due to large positive $U$ of Cu $3d$ electrons, an
antiferromagnetic ground state of undoped cuprates was not predicted by the
LDA \cite{Pick}. If, on the other hand, negative $U$ is due to strong
electron--lattice coupling \cite{Rice}, the LDA should quantitatively
explain the observed instabilities. (It is likely that the nesting idea is
not very convincing as we discussed in Sec.II of this paper. The breathing
instability, for example, has not been found in La$_2$CuO$_4$ where the same
logic is valid.)

Several total energy LDA calculations of the structural phase diagram for
pure BaBiO$_3$ exist in the literature \cite{Liecht1,Zeyher,Blaha,Liecht2}
and the results seem to be inconsistent. Pseudopotential calculations \cite
{Zeyher} predicted rotational but not a breathing instability. In contrast
to that, pure breathing distortion $b\simeq 0.06\AA $ with large energy
lowering $\Delta E_b\simeq $ -50 meV/(1$\times $cell) has been found \cite
{Blaha} by the linear--augmented--plane--wave (LAPW) calculation. A third
calculation \cite{Liecht1} based on the full--potential LMTO method \cite{MM}
gives here nearly zero $\Delta E_b\sim -$0.7 meV/(1$\times $cell) with the
equilibrium value $b\sim 0.03\AA $ for pure breathing. Using the same method 
\cite{MM}, the authors of Ref. \onlinecite{Liecht2} calculate $\Delta E_b=-$%
20 meV/(1$\times $cell) and $b=0.07\AA .\,$ The calculations of the total
energy $E_{tot}$ as a function of combined tilting plus breathing ($tb)$
distortions have also been reported. The first such calculation \cite
{Liecht1} predicts $\Delta E_{tb}=$-14 meV/(1$\times $cell) relative to
rotational energy minimum, $t=8.5^{\circ }$ and $b=0.055\AA $. The second
calculation \cite{Liecht2} gives $\Delta E_{tb}=$-40 meV/(1$\times $cell), $%
t=9.6^{\circ }$ and $b=0.11\AA $. Note that the experimental values found
for $T$=150K are: $t_{exp}=11.2^{\circ }$ and $b_{exp}=0.085\AA .$ A less
rigorous potential--induced--breathing model predicts \cite{PIB,Mazin} these
properties with similar accuracy. In addition, the reported \cite{Liecht1}
frequency $\omega _b$=46 meV of the breathing mode is too low comparing to
the measured \cite{NEUT,Braden} values $\sim 70$ meV.

These large discrepancies in predicting structural properties of BaBiO$_3$
seem to contradict with the accuracy of LDA which standardly is of the order
a few per cent. This could already signalise that the ground state of this
semiconductor is not completely captured within mean--field LDA solution.
Also, another possible reason for the obtained discrepancies could be due to
unusual sensitivity of the calculated properties to the computational
details.

We have performed our own studies of the structural phase diagram for pure
BaBiO$_3$ based on the highly--precise full--potential LMTO method \cite
{FPLMTO}. We have indeed found that there is a sensitivity of the final
results to the details of the calculations, and we will discuss this due
course. However, we have also found that serious errors are introduced by
the LDA. Our general set--up for the calculations is the same as was used
for the doped compound Ba$_{0.6}$K$_{0.4}$BiO$_3$. It has been described at
the beginning of Sec. II of this paper.

As a first step we check an equilibrium cell volume. The value of $%
V/V_{exp}=0.998$ is found and the calculated bulk modulus is equal to 1.29
Mbar. These values essentially depend on the treatment of the semicore
states. Especially, treating deep lying $5d$ Bi states in the main valence
panel, i.e. allowing their full hybridisation with the valence states, is
found to be crucial for the appearance of the total energy minimum itself.
Such sensitivity is due to the fact that the ground state potential $V({\bf r%
})$ is highly not spherical. In the Bi--O directions, the potential exhibits
a pronounced minimum ($\sim 0.5$ Ry down its average value at the Bi
MT--sphere), and the average kinetic energy of Bi $5d$ orbitals in the
interstitial region is nearly zero. As a consequence, the character of Bi $%
5d $ electrons appears at O atoms. In the Bi--Ba directions the potential
exhibits large maximum ($\sim 0.5$ Ry up its average value at the Bi
MT--sphere), and the kinetic energy here is large negative. Kinetic energy
variation about 1 Ry leads to necessity to use multiple $\kappa $ LMTO basis
even for semicore states.

\begin{figure}[bt]
\begin{minipage}[t]{3.25in}
\begin{center}
\epsfig{file=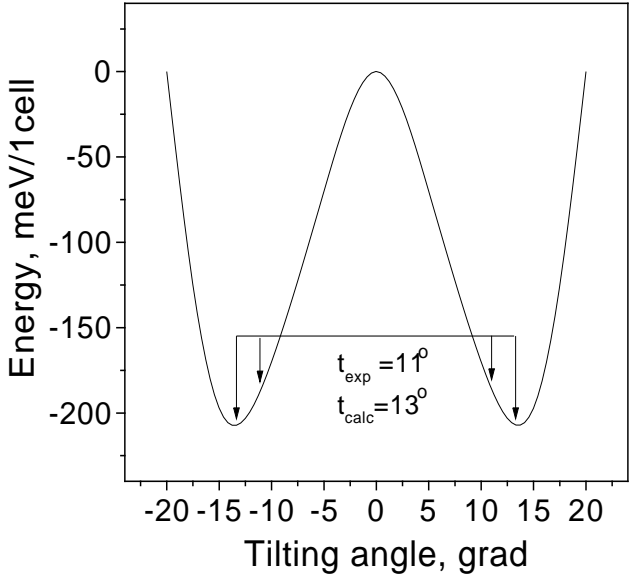,height=3.0in,width=\linewidth,clip=}
\end{center}
\caption{Frozen--phonon calculation of the total energy (meV/1$\times $cell)
as a function of the tilting angle for undoped BaBiO$_3$.}
\label{tilt1}
\end{minipage}
\end{figure}

We now report our results for the calculated tilting instability. The unit
cell is doubled in this calculation according to the $R$--point of the cubic
phase. The rotations of the octahedra are performed along the direction
(1,1,0) as in the experiment \cite{STRUC}. The calculated $E_{tot}$ versus
tilting angle is shown in Fig. \ref{tilt1}. A pronounced minimum
corresponding to the angle $t=13^{\circ }$ is found by our calculation, and
large energy lowering $\Delta E_t=-$200 meV/(1$\times $cell) is predicted.
Note that for the doped compound, the calculated energy lowering becomes
only -10 meV/(1$\times $cell) as can be seen from Fig. \ref{tilt}. The
tilting angle calculated by us agrees with the experiment \cite{STRUC}
within the same accuracy as previously reported \cite{Liecht1,Liecht2}.
However, our calculated $\Delta E_t$ disagrees with $\Delta E_t$=$-$24 meV/(1%
$\times $cell) of Ref. \cite{Liecht1} and with $\Delta E_t$=$-$60 meV/(1$%
\times $cell) of Ref. \cite{Liecht2}. To check the sensitivity of our values
to the treatment of semicore states, we have performed the calculations by
placing Bi $5d$ orbitals into a separate energy panel. This indeed has an
effect in the equilibrium tilting angle equal now $9^{\circ }$ and $\Delta
E_t$=$-$50 meV/(1$\times $cell). Four times change in the latter value looks
very unusual and again shows the importance of proper handling with the
semicore. We have also tried to increase the number of ${\bf k}$--points in
the integration over the BZ but this hardly has any effect to the final
results.

\begin{figure}[bt]
\begin{minipage}[t]{3.25in}
\begin{center}
\epsfig{file=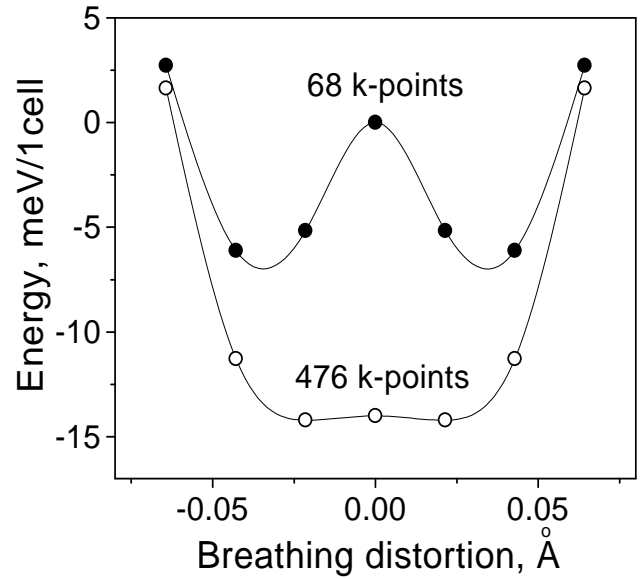,height=3.0in,width=\linewidth,clip=}
\end{center}
\caption{Frozen--phonon calculation of the total energy (meV/1$\times $cell)
as a function of the breathing distortion ($\AA$) for undoped BaBiO$_3$.
The upper curve is obtained using 68 ${\bf k}$--points for the integration
over $\frac {1}{4}$BZ of the monoclinic lattice and the lower curve corresponds
to 476 ${\bf k}$--points.}
\label{breath1}
\end{minipage}
\end{figure}

We now discuss our studies for the breathing distortions. The dimerisations
of the octahedra are performed for a set of tilting angles from 0 to 15$%
^{\circ }$ in order to search for a global total--energy minimum at the $bt$
plane$.$ The crystalline structure consists of two formula units and has now
a monoclinic symmetry. Our calculations do not predict pure breathing
distortions ($t=0^{\circ })$ leaving cubic structure stable against this
perturbation. However, for $t=0^{\circ }$ our total energy displays very
flat highly anharmonic potential well assuming closeness to the instability.
When tilting appears, the potential well starts to exhibit a double minimum
indicating weak but non--zero breathing distortion. We have found that
global total--energy minimum occurs when $t=13^{\circ }$ The calculated $%
E_{tot}$ versus $b$ for this angle is shown in Fig. \ref{breath1}. A very
shallow minimum can be seen from this figure located at $b\sim 0.04\AA .$
The corresponding energy lowering $\Delta E_b$ is only equal to -7 meV/(1$%
\times $cell) relative to the structure which is purely tilted. These
results are in contradictions with the experimental findings \cite{STRUC}
which give $b_{exp}=0.085\AA .$ Moreover such low $\Delta E_b$ gives us the
breathing phonon frequency nearly equal to zero in contrast to the measured
values \cite{NEUT,Braden} of the order $70$ meV. This indicates that real
energy lowering is much larger than we evaluate.

Furthermore, the LDA\ one--electron spectrum calculated at our minimum still
corresponds to metallic ground state. We have found that the energy gap $E_g$
opens at larger values of $b\sim 0.07\AA $. At $b=b_{exp}=0.085\AA $ the
calculated minimal gap is indirect and occurs between the points $X$ and $L$
(see also Ref. \onlinecite{Liecht1}). Its value is approximately equal to
0.1 eV. The minimal direct gap is found at the point $L$ and is about 1 eV.
These values are lower than the corresponding experimental values \cite
{Uchida} most likely due to systematic underestimation of gaps by the LDA..
The measured transport activation gap is 0.24 eV. However, it does not show
up in photoconductivity, optical absorption or photo-acoustic measurements.
Sometimes this is interpreted \cite{Taraphder} as bosonic bound state of two
electrons due to negative electronic $U$. Therefore it is unclear whether
transport activation gap can be related to the minimal indirect gap of our
calculation. The gap seen from the optical measurements is about 2 eV which
is substantially larger than our 1 eV direct gap. We have however not
performed the calculations of the optical properties, therefore an exact
comparison between the theory and experiment is not currently possible.

Several test calculations have been made to check our results. This mostly
concerns the breathing instability, the energy bands were found to be
insensitive to the computational details. First, we place Bi $5d$ orbitals
into a separate energy panel. This has an interesting effect that the
breathing become more pronounced, the calculated value of $b=$ $0.065\AA $
becomes much closer to the experiment, and the energy lowering is now -17
meV/(1$\times $cell) relative to the purely tilted structure. This result,
however, should be considered as artificial, since separate treatment of Bi $%
5d$ states does not allow us to reproduce the equilibrium cell volume. We
second tried to analyse the convergency with respect to the multiple $\kappa 
$ LMTO\ basis. We have used up to 5 $\kappa $ basis functions for
representing valence wave functions but this practically does not affect the
final results. Adding higher lying Ba $6s$ and Bi $6d$ orbitals has no
effect either. This indicates that the basis set described above (see Sec.
II) is sufficiently complete. We third try to investigate the relativistic
effects due to heavy Bi atoms. Inclusion of spin--orbit coupling along with
the scalar relativistic terms was found to have a negligible effect at the
calculated equilibrium structure. Fourth, we increased the number of ${\bf k}
$--points to 476 per $\frac 14$th BZ of the monoclinic lattice. This has an
effect of lowering total energy for metallic states (i.e. when $b\leq $ $%
0.07\AA )$ and practically does not change the values of $E_{tot}$ for
semiconducting energy bands. As a result, total energy as a function of $b$
becomes even more shallow (see Fig. \ref{breath1}) and predicted breathing
at the equilibrium is extremely small.

Based on our findings we conclude that {\em the breathing distortions are
seriously underestimated (ideally absent) in the LDA}, {\em the predicted
ground state is metallic and, therefore, the charge--density-wave
instability is not correctly described.} This situation is analogous to that
with HTSC cuprates where antiferromagnetic ground state was also not found
within LDA \cite{Pick}. (It therefore seems that LDA may, generally, have
problems with both spin--density waves and charge--density waves.) We also
think that due to either improper handling with Bi $5d$ orbitals or
approximate treatment of the full--potential terms in the LMTO method of
Ref. \onlinecite{MM}, previous calculations \cite{Liecht1,Liecht2} did not
converge to the true LDA ground state.

Following to the Hohenberg--Kohn theorem \cite{LDA} we know that more proper
treatment of exchange--correlation effects beyond LDA should in principle
reproduce semiconducting ground state of BaBiO$_3$ and the correct values
for breathing distortions. At the same time, the energy gap will not
necessarily be reproduced since it is not a ground state property of a
single system. However, in such systems as BaBiO$_3$ the energy gap is
directly related to the charge disproportionation between two Bi atoms since
the splitting between occupied and empty Bi $6s$ levels is proportional to
the occupancies of these orbitals. The latter is related to the charge
density distribution which is a ground state property. It is therefore seems
that until the correct theory will not reproduce the energy gap values, the
correct breathing distortions will not be obtained. Speculating on this, (i)
we do not see how the LDA can describe the correct ground state and, at the
same time, strongly underestimate the gap value, (ii) we think the exact
density functional theory would describe both the ground state and the
energy gap in this system.

It worth to mention in this context an example with antiferromagnetic oxides
like NiO. It is known that LDA calculations underestimate both the magnetic
moment and the energy gap value in this compound. It is also clear that the
magnetic moment is given by the occupancies of $3d_{x^2-y^2}$ and $3d_{z^2}$
orbitals of Ni, and the same factor defines the splitting between occupied
and empty states which is directly related to the energy gap. A so--called
LDA plus $U$ method \cite{LDA+U} provides more proper treatment of the
systems with strong electronic correlations. It has been shown \cite{LDA+U}
that this method predicts correct ground state for many Mott--Hubbard
insulators and, at the same time, gives more accurate values for the energy
gaps comparing to the LDA. Also, similar improvements are obtained with the
use of self--interaction corrected density functional \cite{SIC}. If we
would interpret these results as using a better energy functional, then the
situation with NiO should be completely analogous to BaBiO$_3$ with the
exception that not the spin transfer but the charge transfer defines the
properties of this system.

These conclusions may bring an attention to the existence of intra atomic
correlations of Bi $6s$ electrons not captured by the LDA. It does not
necessarily follows from our calculations that the Coulomb $U$ is negative
in these systems. What only follows from our calculations is that {\em there
exists a correction term to the LDA in which the parameter responsible for
the attraction of two electrons at the Bi sites is negative}. To illustrate
this, let us restrict ourselves for the moment by only Bi $6s$ orbitals. The
correction energy for the doubled unit cell Ba$_2$Bi$^{4+q}$Bi$^{4-q}$O$_6$
can be represented in the form (see Appendix)

\begin{equation}
\Delta E_{corr}=\frac 12\sum_{i=Bi1,Bi2}\Delta U_{eff}(n_i-\bar n)^2=\Delta
U_{eff}q^2  \label{b1}
\end{equation}
where $n_i\,$ is the occupancy of the Bi1=Bi$^{4+q}$ or Bi2=Bi$^{4-q}$ $6s$
state and $\bar n=(n_{Bi1}+n_{Bi2})/2.$ The charge disproportionation
parameter $q=(n_{Bi1}-n_{Bi2})/2.$ When $\Delta U_{eff}<0$, the correction (%
\ref{b1}) to the LDA total energy is negative for any non--zero $q$, and
therefore the charge disproportionate state becomes favourable. This
construction is purely heuristic and is built on the analogy to the LDA+U
density functional \cite{LDA+U}. (A better name here could be ''LDA {\em %
minus} U''.) In order to avoid double counting effects we interpret $\Delta
U_{eff}$ as $U-U_{LDA}$, where $U_{LDA}\,$ is a part of the on--site
interaction taken into account in the LDA. (In this way, when the LDA is
adequate for describing the correlations, the correction term becomes
automatically zero.) Therefore, from our calculations follows that the LDA
overestimates Coulomb $U$ for Bi $6s$ electrons, and the difference $\Delta
U_{eff}$ is less than zero. Adding $\Delta E_{corr}$ to the LDA functional
will obviously result in obtaining the charge disproportionation.

Unfortunately this simplified illustrational model has several drawbacks.
First, the antibonding band crossing the Fermi level (see Fig. \ref{bands})
consists not only of the Bi $6s$ electrons but also has a substantial
character of the O $2p$ and Bi $6p$ electrons. Therefore, $\Delta U_{eff}$
in Eq. (\ref{b1}) should be interpreted as the on--site interaction between
the correspondingly constructed Wanier functions. Second, since these Wanier
functions are long ranged, {\em intersite} Coulomb interaction parameters
must be introduced into the expression (\ref{b1}). (In fact, by Fourier
transforming the antibonding band at $E_F$ we evaluate the range of Wanier
states to be 4 lattice constants of the cubic phase.) Another possibility is
to consider a multiband model involving charge fluctuations and Coulomb
interactions between Bi $6s$, O $2p$ and Bi $6p$ states.

The Coulomb interaction parameters can, in principle, be calculated using
constrained density functional method \cite{CLDA1,CLDA2}. However, the
number of parameters required for our system seems to be much larger than
just one number $\Delta U_{eff}$. This gives a lot of extra freedom and
complicates their finding. Despite this difficulty, some progress has
already been made in this direction \cite{Vielsack}. We also plan to
investigate the parameters required for the extended multiband Hubbard model
using constrained LDA calculations. This question will be addressed in the
future publications.

In the following, we take the simplified model (\ref{b1}) to illustrate
which features will bring the inclusion of the correction energy to the LDA
functional. Using a variational principle to minimise our LDA--$U$
expression for the total energy $E_{LDA}+\Delta E_{corr}$ leads to solving
the single--particle equations with the potential $V_{LDA}+\Delta V_{corr}^i,
$ where $\Delta V_{corr}^i=\Delta U_{eff}(n_i-\bar n)=\Delta U_{eff}q$ is
the contribution which has an orbital dependence. It is clear that the value 
$2\Delta V_{corr}^i$ determines the effective splitting between two
non--equivalent Bi $6s$ levels. If our LDA calculation gives this splitting
nearly equal to zero (unless we set the breathing distortion to the
experimental one), then its real value should be of the order of optical gap
experimentally determined as 2 eV. (Simple connection between the charge
disproportionation and the gap value is clearly seen from here, and we again
state, that we do not see the way how the LDA can obtain the correct
breathing and at the same time seriously underestimate $E_g.$)

We perform the calculations involving $\Delta V_{corr}^i\,$ using our
full--potential LMTO method \cite{FPLMTO}. The calculations are analogous to
the constrained LDA calculations described in Ref. \onlinecite{CLDA2}. Using
projector operator technique we define the corrections to the Bi $6s$
diagonal matrix elements of the LMTO Hamiltonian. The projection is simply
taken to the $l=0$ spherical harmonic inside the MT sphere of the Bi site
and zero everywhere else. According to our model, the difference between two
non--equivalent Bi $6s$ occupation numbers should be associated with the
charge disproportionation $2q.$ If we now fix $\Delta U_{eff}$ to some
value, the self--consistent procedure will define this charge
disproportionation, and the correction $\Delta E_{corr}$ to the LDA total
energy can be estimated. In this way, we can find a new equilibrium
structure and compare that with the experiment. Choosing such $\Delta U_{eff}
$ to reproduce the experimental breathing distortion can give some insight
on this value.

We have performed a set of self--consistent calculations of the total energy 
$E_{LDA}+\Delta E_{corr}$ as a function of breathing distortion for
different values of $\Delta U_{eff}$. We have indeed found that for $\Delta
U_{eff}$ of the order minus 10 eV, the experimental breathing distortion is
fairly well reproduced and the correct breathing phonon frequency is
obtained. Also, the energy gap becomes much closer to the experiment. It is
interesting to note that the form (\ref{b1}) of the functional assumes
broken symmetry, i.e. {\em the existence of the charge disproportionation
regardless the presence of breathing distortions}. As a consequence, the
double--potential well does not exhibit a smooth behaviour at $b$=0 and has a
kink there.

The value of $\Delta U_{eff}=$--10 eV required to describe the experimental
structure seems to be unphysically large. This is mainly due to small
self--consistent values of $q$ of the order 0.1 electrons which are obtained
in our calculations. (Even smaller values $\sim 0.01e$ are obtained for the
charge transfer between two Bi MT--spheres because of the screening by the
tails of O 2p orbitals). In fact, if we take a simple relationship $2q\Delta
U_{eff}=2\Delta V_{corr}^i\sim E_g=2\,$ eV, the same $\Delta U_{eff}=$--10
eV is obtained. This again shows that our logic is valid. As we discussed
above, a better way would be to interpret $2q$ as a charge transfer between
two Wanier states centred in the sites of cubic lattice with two
non--equivalent Bi atoms. (We associate these Wanier states with the Bi
atoms, but it is clear that they represent a mixture of a whole set of
orbitals.) Indeed, our tight--binding calculations for a one--band model
with long--range interactions (fitted to describe the dispersion of the band
crossing the Fermi level) show that introducing a small splitting between
the effective Bi levels gives the charge transfer $q$ of the order $1$ $e$.
(The chemical interpretation of the charge disproportionation 2Bi$%
^{4+}\rightarrow $Bi$^{3+}$+Bi$^{5+}$ is thus valid for the Wanier functions
and is, of course, wrong for the real charge densities.) Unfortunately,
implementing the Wanier representation in our LDA-$U$ functional is not
straightforward and we postpone this for the future work. However, a rough
estimate of $\Delta U_{eff}$ operative between the Wanier states can be
found by setting $q$ to 1 in the relationship $2q\Delta U_{eff}\sim 2\,$ eV.
This will result in more realistic value of $\Delta U_{eff}$=--1 eV. Such
values are expectable from the errors introduced by the LDA, for example,
due to self--interaction effects.

This discussion shows that the proper treatment of the parameters of the
model (\ref{b1}) is required. It is also clear that inclusion of the {\em %
intersite} Coulomb interactions is necessary to obtain a quantitative
description of the ground state in BaBiO$_3.$ The above calculations cannot
be considered as estimates of $U.$ Therefore we would await of making the
conclusion that negative on--site $U$ of Bi $6s$ orbitals is of the
electronic origin until we will not evaluate the errors introducing by the
LDA. In the above discussion we were trying to argue that the insulating
behaviour and breathing distortions in this system are directly related to
the charge disproportionation, and the LDA must fail in describing this
ground state already because it is not able to reproduce the energy gap. On
the other hand, the LDA$+U$ like techniques were seen to be very perspective
for performing accurate calculations, and we will try to address this
subject in the future work.

\begin{center}
{\bf V. CONCLUSION.}
\end{center}

In the present paper we have reported our density functional LDA studies of
the compound Ba$_{1-x}$K$_x$BiO$_3.$ For its superconducting phase ($x$=0.4)
we have performed the calculations of full wave--vector--dependent lattice
dynamical properties and the electron--phonon interactions. These
calculations were based on recently developed linear--response approach
implemented in the framework of the full--potential LMTO method. The
following conclusions are made on the basis of our studies: The calculated
phonon dispersion curves along major symmetry directions agree reasonably
well with the results of the neutron--scattering experiments. Some
discrepancies have been found to occur in reproducing the bond--stretching
oxygen modes. They were attributed to the virtual crystal approximation used
for the treatment of potassium doping. It was found that the
bond--stretching modes have large coupling to the electrons. Especially the
breathing modes at the R and M points of the cubic Brillouin zone have large
coupling equal to 0.3. Relatively strong coupling is also predicted for the
bond--bending oxygen modes. However, our calculated averaged value of $%
\lambda $ was found to be 0.29 which is too small to account for the
high--temperature superconductivity in the doped barium bismuthates. These
results were supported by the calculated transport properties such, e.g., as
electron--phonon limited electrical resistivity.

Our own and previous frozen--phonon calculations predicted highly anharmonic
double--potential well behaviour for the tilting of oxygen octahedra
corresponding to the point R. We have performed detailed studies on the
influence of anharmonicity corrections to the electron--phonon coupling. By
neglecting the processes of phonon--phonon interactions and utilising
zero--temperature treatment of Hui and Allen \cite{Hui}, we have worked out
the formula for anharmonic contribution to $\lambda $ up to the
second--order with respect to the displacements. It was found that
anharmonic $\lambda $ can be not small if large ionic excursions take place
as in the case of the tilting vibrations. Using the full--potential LMTO
method and frozen--phonon approach, we have estimated that contribution
equal to 0.04. Our total $\lambda $ is thus 0.33. We concluded that while
not negligible, anharmonicity corrections due to tiling modes do not help to
explain the superconductivity in the bismuthates, and therefore, its origin
still remains open and intriguing problem.

As a final issue, we have done the calculations of the structural phase
diagram for the undoped parent compound BaBiO$_3.$ A low--temperature
experimental structure consists of combined breathing plus tilting
distortions. In agreement with the previous conclusions, we have found that
tilting distortions are reasonably well reproduced by the LDA calculation.
However, we have also found that the breathing distortions are seriously
underestimated (if not absent) within the LDA. This contradicts with the
previous findings most likely due to improper handling with the semicore
states. The underestimation of breathing leads to the predicted ground state
which is metallic while it should be insulating of the charge--density--wave
type. This situation closely resembles antiferromagnetism of HTSC cuprates
which was also not predicted by the LDA. Using a simple correctional scheme
in the spirit of the LDA$+U$ method we have tried to argue that the problem
of breathing is most likely due to underestimation of energy gap by the LDA.
While we cannot make any definite conclusions on whether negative $U$ in the
bismuthates is of the electronic origin, the failure of the LDA indicates
that the electron correlations of the Bi $6s$ electrons are not properly
treated. Since, LDA calculations contain all electron--lattice coupling
effects, we conclude that in order to recover an insulating state, an
existence of the correction to the LDA with some negative $\Delta U_{eff}$
of the electronic origin must be assumed. We believe that LDA$\pm U$ like
approaches will help in further understanding of the physics in these
systems from density functional point of view.

\begin{center}
{\bf ACKNOWLEDGEMENTS}
\end{center}

The authors are indebted to O. K. Andersen, O. Gunnarsson, O. Jepsen, A.
Liechtenstein, E. G. Maksimov, Y. Yacoby,  and R. Zeyher, 
for many helpful discussions. The work was
performed under the Research Contract No. I 0457-222.07/95 
supported by the German--Israel
Foundation for scientific research and development.

\begin{center}
{\bf Appendix.}
\end{center}

The form of the correction can be understood using the following arguments:
Suppose the LDA total energy $E_{LDA}$ gives non--degenerate ground state
described by the density $\rho _0$ as is the case of predicting BaBiO$_3$
without breathing distortions. Let us expand $E_{LDA}(\rho )$ around its
minimum. Due to extremal property, this expansion starts from the
second--order variations: 
\[
E_{LDA}(\rho )=E_{LDA}(\rho _0)+\frac 12\int \frac{\delta ^{(2)}E_{LDA}}{%
\delta \rho \delta \rho }(\rho -\rho _0)(\rho -\rho _0)+... 
\]
If the predicted ground state is wrong, then the first candidate which can
take responsibility for this is the second--order derivative. The value of $%
\delta ^{(2)}E_{LDA}/\delta \rho \delta \rho $ at $\rho =\rho _0$ is
positively defined since $E_{LDA}(\rho )$ has a minimum here. On the other
hand, the true density functional at $\rho _0$ would indeed have a local
maximum or, more generally, a saddle point, since the true ground state is
double degenerate corresponding to either breathing--in and breathing--out
distortions or vice versa. Therefore, $\delta ^{(2)}E_{true}/\delta \rho
\delta \rho $ should be negatively defined at $\rho =\rho _0$. (We assume
that the first--order variation $\delta ^{(1)}E_{true}/\delta \rho $ is
equal to zero at $\rho =\rho _0$ due to symmetry of the saddle point.) It is
thus tempting to construct a density functional with the corrected
second--order variation, i.e. 
\begin{eqnarray}
&&E_{corr}(\rho )=E_{LDA}(\rho )+  \nonumber \\
&&\frac 12\int \left( \frac{\delta ^{(2)}E_{true}}{\delta \rho \delta \rho }-%
\frac{\delta ^{(2)}E_{LDA}}{\delta \rho \delta \rho }\right) (\rho -\rho
_0)(\rho -\rho _0)  \nonumber
\end{eqnarray}
Assuming that kinetic energies of both $E_{true}$ and $E_{LDA}$ functionals
are the same, the difference between the second--order derivatives is
described by effective Coulomb interactions. This leads to the form (\ref{b1}%
) of the correction.

\end{multicols}


\begin{references}
\bibitem{comm1}  This definition is different from that we have used in Ref. %
\onlinecite{EPI} (no factor of 2 appeared in the denominator of the matrix
element). However it is now more easy to make a comparison with the matrix
element over the phonon states (\ref{as9}) discussed in Sec. III.
\bibitem{comm3}  It should be noted that Eq. \ref{a19} does not work when $%
\omega _1\rightarrow 0$ (a case of two separated potential wells)\ since the
starting expression (\ref{a1}) is only valid when energy levels $\epsilon _n$
are not degenerate and well separated from each other. The corresponding
criterium of applicability of (\ref{a1}) is that the distance between levels 
$\epsilon _n$ should be much larger than applied perturbation described by
electron--phonon interactions. In the case close to degeneracy, another
formula should be used which is based on the perturbation theory allowing
degenerate states.
\bibitem{comm2}  It is not generally true that $\langle \xi |\delta
^{(2)}V|\xi \rangle \,\,$ is equal to the second--order change in the
eigenvalue $\delta ^{(2)}E_\xi $ because, following to standard perturbation
theory, there also is a contribution $\sum_\zeta ^{^{\prime }}|\langle \zeta
|\delta ^{(1)}V|\xi \rangle |^2/(E_\xi -E_\zeta ).$ of the same second
order. However, for the tilting mode, all contributions to $\lambda $
connected with $\delta ^{(1)}V$ vanish.
\end{references}
\end{document}